\documentclass[twocolumn]{aastex61}
\usepackage{natbib}
\usepackage{graphicx}

\submitjournal{ApJ}

\shorttitle{edge-on HUDS}
\shortauthors{He et al.}

\begin{document}
%\begin{CJK*}{UTF8}{kai}
\nocite{*}
\title{Edge-on H{\sc{i}}-bearing ultra diffuse galaxy candidates in the 40\% ALFALFA catalog}

\correspondingauthor{Hong Wu, Wei Du, Min He}
\email{Email: hwu@bao.ac.cn, wdu@bao.ac.cn, hemin13@live.com}

\author{Min He}
\affil{Key Laboratory of Optical Astronomy,
       National Astronomical Observatories,
       Chinese Academy of Sciences,
       20A Datun Road, Chaoyang District,
       Beijing 100101, China}

\affiliation{School of Astronomy and Space Science,
             University of Chinese Academy of Sciences,
             19A Yuquan Road, Shijingshan District,
             Beijing, 100049, China}

\author{Hong Wu}
\affiliation{Key Laboratory of Optical Astronomy,
             National Astronomical Observatories,
             Chinese Academy of Sciences,
             20A Datun Road, Chaoyang District,
             Beijing 100101, China}
\affiliation{School of Astronomy and Space Science,
             University of Chinese Academy of Sciences,
             19A Yuquan Road, Shijingshan District,
             Beijing, 100049, China}

\author{Wei Du}
\affiliation{Key Laboratory of Optical Astronomy,
             National Astronomical Observatories,
             Chinese Academy of Sciences,
             20A Datun Road, Chaoyang District,
             Beijing 100101, China}

\author{James Wicker}
\affiliation{Chinese Academy of Sciences,
             20A Datun Road, Chaoyang District,
             Beijing 100101, China}

\author{Pin-song Zhao}
\affiliation{Key Laboratory of Optical Astronomy,
             National Astronomical Observatories,
             Chinese Academy of Sciences,
             20A Datun Road, Chaoyang District,
             Beijing 100101, China}
\affiliation{School of Astronomy and Space Science,
             University of Chinese Academy of Sciences,
             19A Yuquan Road, Shijingshan District,
             Beijing, 100049, China}

\author{Feng-jie Lei}
\affiliation{Key Laboratory of Optical Astronomy,
             National Astronomical Observatories,
             Chinese Academy of Sciences,
             20A Datun Road, Chaoyang District,
             Beijing 100101, China}
\affiliation{School of Astronomy and Space Science,
             University of Chinese Academy of Sciences,
             19A Yuquan Road, Shijingshan District,
             Beijing, 100049, China}

\author{Ji-feng Liu}
\affiliation{Key Laboratory of Optical Astronomy,
             National Astronomical Observatories,
             Chinese Academy of Sciences,
             20A Datun Road, Chaoyang District,
             Beijing 100101, China}
\affiliation{School of Astronomy and Space Science,
             University of Chinese Academy of Sciences,
             19A Yuquan Road, Shijingshan District,
             Beijing, 100049, China}

\begin{abstract}

Ultra-diffuse galaxies (UDGs) are objects which have very extended morphology and faint central surface brightness. Most UDGs are discovered in galaxy clusters and groups, but also some are found in low density environments. The diffuse morphology and faint surface brightness make them difficult to distinguish from the sky background. Several previous works have suggested that at least some UDGs are consistent with exponential surface brightness profiles (S\'{e}rsic $n \sim 1$). The surface brightness of exponential disks is enhanced in edge-on systems, so searching for edge-on systems may be an efficient way to select UDGs. In this paper, we focus on searching for edge-on H{\sc{i}}-bearing ultra-diffuse sources (HUDS) from the 40\% ALFALFA catalog, based on SDSS $g$- and $r$-band images. After correcting the observed central surface brightness to a face-on perspective, we discover 11 edge-on HUDS candidates. All these newly discovered HUDS candidates are blue and H{\sc{i}}-bearing, similar to other HUDS in 70\% ALFALFA catalog, and different from UDGs in clusters.\\

\end{abstract}


\section{Introduction} \label{sec:introduction}

\defcitealias{Leisman..2017ApJ...842..133L}{L17}
\defcitealias{van.Dokkum..2015ApJ...798L..45V}{V15a}
\defcitealias{van.Dokkum..2015ApJ...804L..26V}{V15b}

Ultra-diffuse galaxies (hereafter UDGs) were first discovered in the Coma cluster by \citet[][hereafter V15a and V15b]{van.Dokkum..2015ApJ...798L..45V, van.Dokkum..2015ApJ...804L..26V}, and have very faint central surface brightness $\mu_{0,g}\geqslant 24\ mag\ arcsec^{-2}$ and very large effective radius $r_{e}\geqslant 1.5\ kpc$, even as large as that of the Milky Way. After that, more UDGs were discovered in galaxy clusters. \cite{Koda..2015ApJ...807L...2K} found nearly 1,000 UDGs in the Coma Cluster, \citet{Munoz.2015ApJ.813.L15} and \citet{Mihos..2015ApJ...809L..21M} identified UDGs in the Fornax Cluster, \cite{Yagi..2016ApJ..225..11} located 854 Subaru-UDGs in the Coma Cluster, \cite{Roman..2017MNRAS.468..703R} identified 80 UDGs in the cluster Abell 168, \cite{van.der.Burg..2017A&A.607.A79} found $\sim 2,500$ UDGs in eight nearby MENeaCS clusters, and \cite{Venhola..2017A&A..608..A142} located 9 UDGs in the Fornax cluster.\\

In addition to the UDGs detected in high density environments, there are also many UDGs that have been detected in lower density environments, some group samples \citep[e.g.][]{martinez..2016AJ..151..96M, Merritt..2016ApJ...833..168M, Roman..2017MNRAS.468.4039R, Trujillo..2017ApJ...836..191T, Shi..2017ApJ.846.26, Bennet..2018ApJ...866L..11B, Spekkens..2018ApJ...855...28S}, and a few field sources: H{\sc{i}}-bearing ultra-diffuse sources (HUDS) found in the $70\%$ ALFALFA catalog (hereafter $\alpha .70$) by \citet[][hereafter L17]{Leisman..2017ApJ...842..133L} and two extended dwarf irregular galaxies \citep{Bellazzini.2017MNRAS.467.3751B}, very similar to UDGs.\\

The origins of UDGs are still unclear. \citetalias{van.Dokkum..2015ApJ...798L..45V} suggest that the UDGs are ``failed'' $\sim L_\star$ galaxies. \cite{Amorisco..2016mnras..459..L51} and \cite{Lim..2018ApJ...862...82L} suggest that UDGs are dwarf galaxies. However, after considering the work of both \cite{van.Dokkum..2015ApJ...798L..45V, van.Dokkum..2015ApJ...804L..26V} and \cite{Munoz.2015ApJ.813.L15}, and according to the study of \cite{Zaritsky..2017MNRAS.464L.110Z}, UDGs may have multiple origins. They may be both ``failed'' $\sim L_\star$ galaxies and dwarf galaxies.\\

UDGs may also have various formation mechanisms, such as being formed by collision \citep{Baushev..2018NA..60..69} or being reproduced by tidal stripping of dwarf galaxies within clusters \citep{Carleton..2018arXiv}. The formation mechanism of gas-rich UDGs, which have bluer colors, is feedback-driven expansion \citep{Di.Cintio.2017MNRAS.466L.1D, Papastergis.2017A&A.601.L10, Chan..2018MNRAS.478..906C}. These blue UDGs can be detected by H{\sc{i}} surveys, such as the Arecibo Legacy Fast ALFA (ALFALFA) extragalactic H{\sc{i}} survey \citep[][\citetalias{Leisman..2017ApJ...842..133L}]{Haynes..2011AJ..142..170}.\\

Compared with optically selected UDGs, H{\sc{i}}-selected UDGs tend to be dominated by gas-rich UDGs, which have much bluer color and prefer to inhabit much lower density environments. Though H{\sc{i}} surveys are powerful tools for detecting field UDGs since they directly measure redshift, they are biased against gas-poor UDGs in the field. For example, \citet{Jones..2018A&A...614A..21J} use the Santa Cruz Semi-analytic model \citep{Somerville.2015MNRAS.453.4337S} to predict the population of field UDGs. This model produces nearly 10 times more objects than they observe in the HUDS population.\\

Between the diffuse blue field sources and red UDGs, it seems that there may be an evolutionary connection. Some literature \citep[e.g.][]{Yozin..2015MNRAS.453.2302Y, Amorisco..2016mnras..459..L51, Rong..2017MNRAS.470.4231R, Di.Cintio.2017MNRAS.466L.1D} suggest that progenitors of red UDGs are the high-spin tail of field dwarf galaxies or dwarf galaxies that are undergoing feedback-driven gas outflows. \cite{Carleton..2018arXiv} apply a semi-analytic model to tidally-stripped UDGs, and the results indicate that for dwarf galaxies which settle in cored halos, the tidal stripping mechanism can reproduce the observed properties of UDGs in clusters. Therefore, blue UDGs may play an important role in the origin of red UDGs.\\

For studying more details about the properties of UDGs, spectral observation is necessary. However, the faint surface brightness makes spectral observation much more difficult. Many previous works have shown that most UDGs are well fitted by S\'{e}rsic model with $n \sim 1$ \citep[e.g. \citetalias{van.Dokkum..2015ApJ...798L..45V}][]{Koda..2015ApJ...807L...2K} or $n<1$ \citep[e.g.][]{Yagi..2016ApJ..225..11, Venhola..2017A&A..608..A142, Roman..2017MNRAS.468..703R, Merritt..2016ApJ...833..168M}, while only a few UDGs are well fitted by $n > 1$, even $n \sim 4$ \citep[e.g.][]{Yagi..2016ApJ..225..11, Lee..2017ApJ...844..157L, Muller..2018A&A...615A.105M}. \citetalias{Leisman..2017ApJ...842..133L} suggests that the profile of HUDS is better fitted by $n = 1$. This may indicate that these galaxies may be exponential-profile galaxies. For this kind galaxy, their edge-on perspective would enhance their surface brightness, making them easier to find and also enabling optical spectral observations. Also, their edge-on direction is suitable for measuring their maximum rotating velocities.\\

In this paper, we use the SDSS DR7 images matched with the 40\% ALFALFA H{\sc{i}} survey to find edge-on HUDS candidates. In Section \ref{sec:data}, we introduce the data we used and the data reduction; In Section \ref{sec:select}, we select the edge-on HUDS candidates after correcting the central surface brightness to face-on orientation; In Section \ref{sec:results}, we compare properties of edge-on HUDS candidates with those of other UDG samples, and present the result of our candidates; In Section \ref{sec:discussion}, we discuss the uncertainty, selection effect and mechanisms of our HUDS candidates; Finally, a summary is given in Section \ref{sec:summary}.\\

\section{Data and data reduction} \label{sec:data}
\subsection{Data}
The Arecibo Legacy Fast ALFA (ALFALFA) extragalactic H{\sc{i}} survey probes the population of local H{\sc{i}}-rich sources over a cosmologically significant volume \citep{Giovanelli..2005AJ.130.2598G, Haynes.2018ApJ.861.49H}. It covers 7000 deg$^2$ of the sky at high Galactic latitude. A catalog based on 40\% of the overall ALFALFA survey sky coverage was released in 2011 (hereafter referred to as $\alpha .40$)\citep{Haynes..2011AJ..142..170}. It consists of 15,855 objects and covers 2800 deg$^2$ of the sky: $07^h30^m < R.A. < 16^h30^m$, $+04\arcdeg  < decl. <  +16\arcdeg $, and $+24\arcdeg  < decl. < +14\arcdeg $ (the ``spring'' range) and $22^h < R.A. < 03^h$, $+14\arcdeg  < decl. < +16\arcdeg $, and $+24\arcdeg  < decl. < +32\arcdeg $ (the ``fall'' range). From its catalog, we can get many useful properties of galaxies, H{\sc{i}} masses, H{\sc{i}} profiles, velocity widths ($W_{50}$), redshifts, and so on. The H{\sc{i}} masses of detected objects are from $10^6M_\sun$ to $10^{10.8}M_\sun$.\\

The Sloan Digital Sky Survey (SDSS)\citep{York..2000AJ.120.1579Y} consists of five bands ($ugriz$), composed of imaging and spectroscopic surveys. It covers 11,663 deg$^2$ of the sky, and has a field that overlaps with ALFALFA. In $\alpha .40$ \citep{Haynes..2011AJ..142..170} there are 12,468 objects that have been matched with their optical counterparts (OCs) in SDSS DR7 \citep{Abazajian.2009ApJS.182.543A}. To compare with previous works, we adopt the $g$- and $r$- band SDSS images for searching and studying edge-on HUDS. For identifying selected objects, we also have checked their optical morphology with the Dark Energy Camera Legacy Survey\footnote{\url{http://legacysurvey.org/decamls/}}, which is much deeper than SDSS and can provide higher quality images.\\

\subsection{Data reduction}
\label{reduction}
The surface brightnesses of UDGs are too faint (nearly three magnitudes fainter than that of dark sky background) to be detected. Accurate background estimation is very important for searching for UDGs, but the Photo Pipeline of SDSS DR7 is not good at searching for low surface brightness objects. It always overestimates the sky background, and underestimates the luminosities and effective radius of galaxies. The average of the underestimation is $0.16$ mag, and the maximum is $0.8$ mag \citep{Du_Wei_2015AJ..149..199, Lauer..2007ApJ..662..808, Liu..2008MNRAS..385..23, Hyde..2009MNRAS..394..1978, He..2013ApJ..773..37}.\\

To eliminate such deviation, we adopt a more adaptive measurement, which has been reported by \cite{Zheng..1999AJ..117..2757}, \cite{Wu..2002AJ..123..1364} and \cite{Du_Wei_2015AJ..149..199}, to estimate the sky background of SDSS fpC-images of 12,468 galaxies using optical-H{\sc{i}} cross matches from \cite{Haynes..2011AJ..142..170}. Firstly, we detect all the objects in a fpC-image by SExtractor \citep{Bertin..1997AA..117..393B} after smoothing the image by a Gaussian Function with the FWHM of 8 pixels; Then we mask these detected objects and derive a sky background image by using both line and column linear fitting method \citep{Zheng..1999AJ..117..2757, Wu..2002AJ..123..1364, Du_Wei_2015AJ..149..199}.\\

Next, we use SExtractor again to perform surface photometry for the target objects after subtracting the sky background from fpC-image. We adopt the AUTO photometry by the Kron flexible elliptical aperture \citep{Bertin..1997AA..117..393B}. The final magnitudes are calibrated by the following formula from the SDSS DR7 website\footnote{\url{http://classic.sdss.org/dr7/algorithms/fluxcal.html}}
\begin{equation}
mag = -2.5\times lg(\frac{counts}{exptimes})-(aa+kk\times airmass).
\end{equation}
Here, `aa' is the zero point of fpC-image and `kk' is the atmosphere extinction coefficient. `aa', `kk' and `airmass' are all given from drField*.fit files of SDSS DR7. `Counts' is measured from AUTO aperture by SExtractor in ADU units, and `exptimes' is the exposure time of SDSS imaging, 53.907456 seconds. Additionally, we correct the measured magnitude for Galactic extinction which is calculated by the dust map of \cite{Schlegel.1998ApJ...500..525S}.\\

\section{The Selection OF HI-bearing Ultra-Diffuse Sources} \label{sec:select}
\subsection{GALFIT fitting}

\begin{figure*}[ht!]
\gridline{\fig{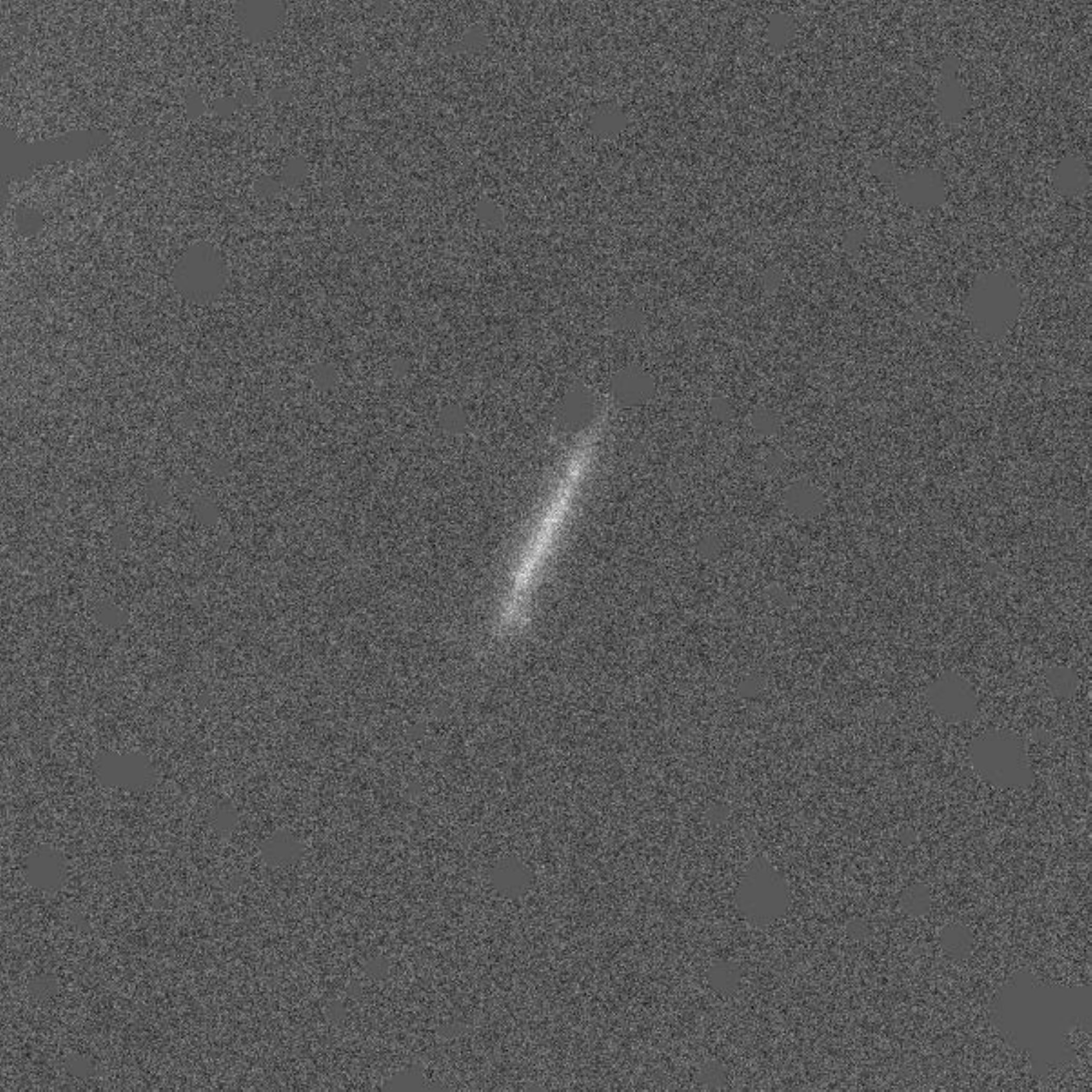}{0.3\textwidth}{fpC-image}
          \fig{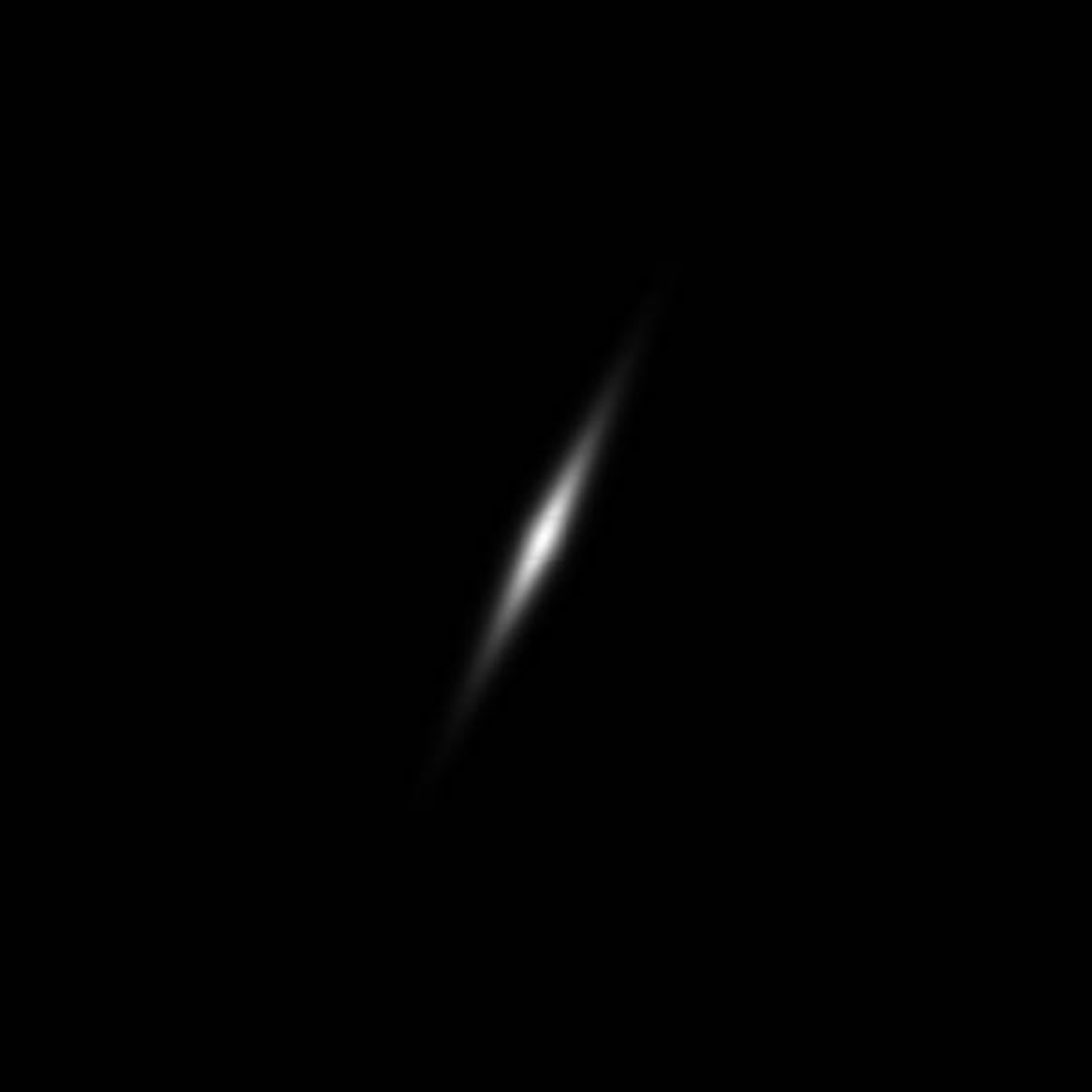}{0.3\textwidth}{edge-on-disk model}
          \fig{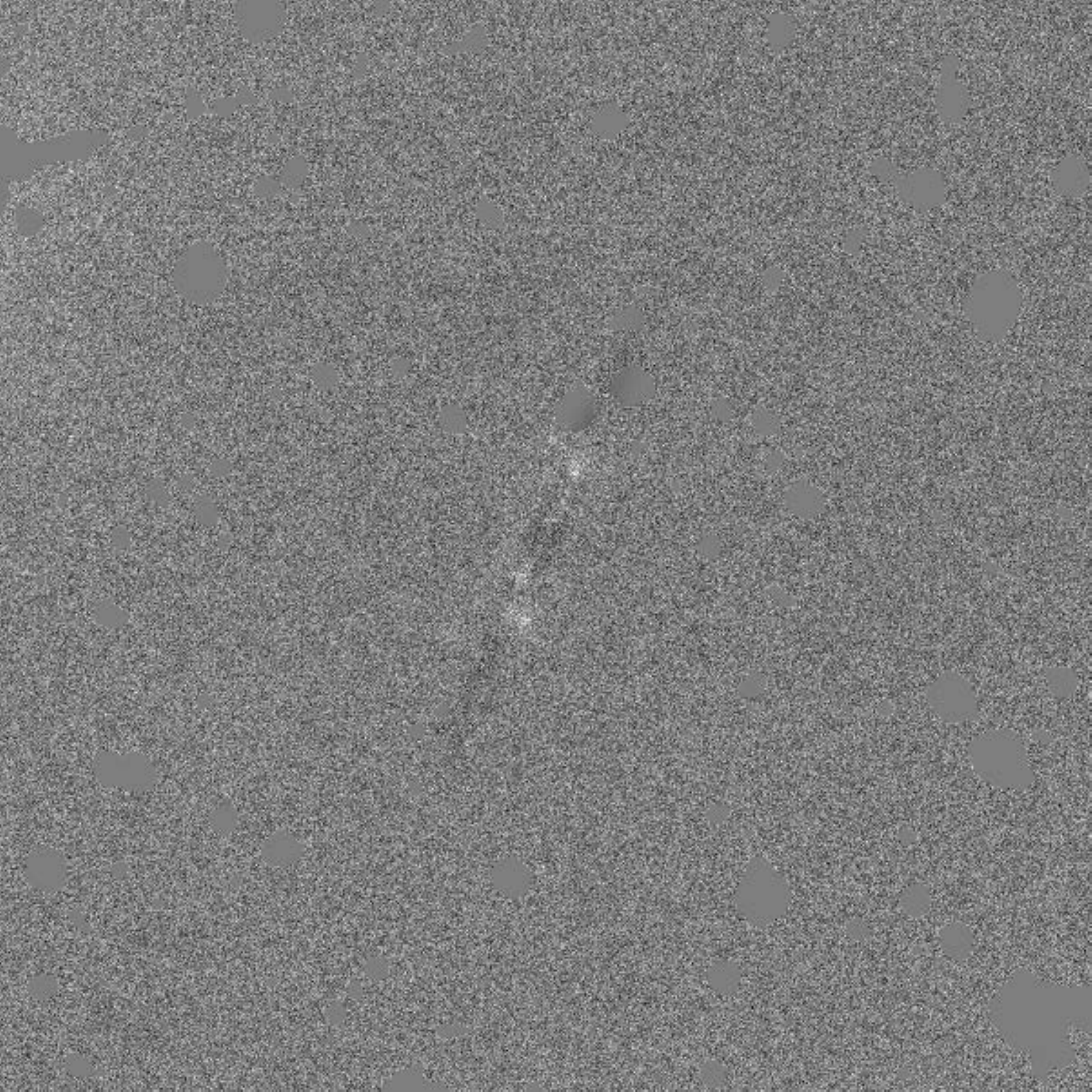}{0.3\textwidth}{residual image}
          }
\caption{Edge-on disk model fitting to the galaxy AGC 202262. These images are the fpC-image after sky subtraction and bright star removal, the model image and the residual image, from left to right respectively.}
\label{fig:galfit}
\end{figure*}

Previous UDG studies \citep[e.g. \citetalias{van.Dokkum..2015ApJ...798L..45V}][]{Koda..2015ApJ...807L...2K, Yagi..2016ApJ..225..11, Venhola..2017A&A..608..A142, Roman..2017MNRAS.468..703R, Merritt..2016ApJ...833..168M} show that the surface brightness profiles of UDGs are best fitted by S\'{e}rsic model $n \sim 1$ or $n < 1$. And \citetalias{Leisman..2017ApJ...842..133L} suggests that $n = 1$ is best for HUDS. So we use the $n=1$ in this work either. Using the exponential profile model by GALFIT \citep{Peng..2010AJ..139..2097}, we can obtain the minor-to-major axis ratio, $b/a$, of each object. In the GALFIT fitting, the PSF image directly achieved from the SDSS is convolved to the galaxy image to correct the PSF smearing.\\

\cite{Du_Wei_2015AJ..149..199} have selected LSBGs with the ratio of $b/a > 0.3$ from the $\alpha .40$ catalog in both the $g$-band and $r$-band. This set excludes the edge-on galaxies. On the contrary, we use all these remaining cases of \cite{Du_Wei_2015AJ..149..199}, whose $b/a \leqslant 0.3$. Then we check images of these galaxies, and reject obviously irregular and interacting galaxies. We also remove some galaxies which are contaminated with nearby bright objects. Finally, 1670 edge-on galaxies remain. These galaxies have symmetrical edge-on disk-like shapes with no visible spiral structures or dust lanes passing through their centers.\\

Then, we mask bright stars around the galaxies in the image, and use GALFIT with edge-on exponential disk model for getting more specific central surface brightness values of these edge-on galaxies. This fitting is sensitive to initial values, such as central surface brightness $\mu_{0,edge}$, scale length $r_{s}$, scale height $h_{s}$ and positional angle $PA$. If these values are far from the best fit, the fitting will easily break down.\\

As we do not know the exact values of these parameters, we give a range of each parameter to fit, and select initial values randomly from the range. We take a test for checking the scatter of fitting results by giving different initial values: the $\mu _{0,edge}$ is a random number in a range of 10 to 25, the $r_{s}$ is in the range from 2 pixels to 1.5 times the disk scale length which is the output of SExtractor in Section \ref{reduction}, the $h_{s}$ is from 1 pixel to half of the input initial value of $r_{s}$, and the $PA$ is the output of SExtractor which is transferred to GALFIT coordinates. We take 100 loops in the test, and there are 34 loops that get good fitting results (a good result means there is not star signs in the result like ``*0.01*''). The standard deviations of these results are 0.000057, 0.001387, 0.000077 and 0.000461 for $\mu _{0,edge}$, $r_{s}$, $h_{s}$ and $PA$ respectively. The output results are convergent. So, once the fitting gets a good result in a loop, we can treat it as the best fitting result we need and continue to fit other galaxies. We display the fitting images of one edge-on galaxy AGC 202262 in Figure \ref{fig:galfit} as an example.\\

\subsection{Central surface brightness dependence on inclination}
However, this central surface brightness is just an observational one, and cannot represent the face-on exponential profile of the system. In fact, both central surface brightnesses are different and have a relation. From \cite{van.der.Kruit..1981A&A..95..105V, Giovanelli..1995AJ....110.1059G}:\\
the face-on disk model is:
\begin{equation}
\tilde{\mu}(R) = \tilde{\mu}_{0,face} e^{-\frac{R}{r_{s}}},
\end{equation}
with $\tilde{\mu}_{0,face}=2h_{s}\rho_{0}$, while \\
the edge-on disk mode is:
\begin{equation}
\tilde{\mu}(R,h) = \tilde{\mu}_{0,edge}(\frac{R}{r_{s}})K_{1}(\frac{R}{r_{s}})sech^2(\frac{h}{h_{s}}),
\end{equation}
with $\tilde{\mu}_{0,edge} = 2r_{s}\rho_{0}$. Here, $\tilde{\mu}$ is surface brightness in luminosity unit, $R$ is the radial distance from the center of the galaxy, and $h$ is the vertical distance from the disk. $\tilde{\mu}_{0}$ is central surface brightness, $\rho_{0}$ is central luminosity density, $r_{s}$ and $h_{s}$ are scale length and scale height of galaxy respectively, and $K_{1}$ is the modified Bessel function. Therefore there is a relationship between the central surface brightness of face-on and edge-on disk galaxies:
\begin{equation}
\label{eq:correct-lumi}
\tilde{\mu}_{0,face}=\tilde{\mu}_{0,edge}\times h_{s}/r_{s}
\end{equation}
and transforming to magnitude unit, this becomes:
\begin{equation}
\label{eq:correct-mu} \mu_{0,face}=\mu_{0,edge}-2.5\times
lg(h_{s}/r_{s}).
\end{equation}
Such central surface brightness of an edge-on galaxy is brighter than that of a face-on galaxy, as scale length is larger than scale height.\\

\begin{figure*}[ht!]
\centering
{\bf \hspace{0.01in} AGC~102276 \hspace{0.26in} AGC~113816 \hspace{0.26in} AGC~198457 \hspace{0.26in} AGC~202262 \hspace{0.26in} AGC~215226 \hspace{0.26in} AGC~219242}\\
\vspace{0.1cm}
\includegraphics[width=0.16\textwidth]{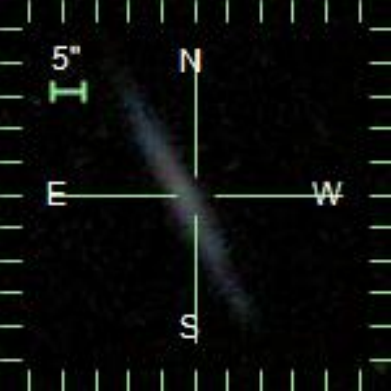}
\includegraphics[width=0.16\textwidth]{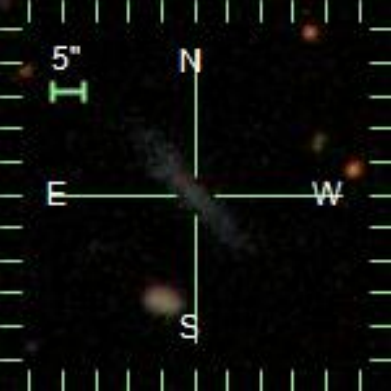}
\includegraphics[width=0.16\textwidth]{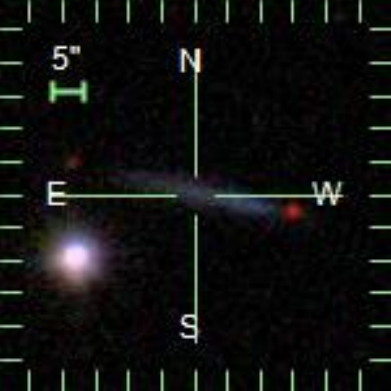}
\includegraphics[width=0.16\textwidth]{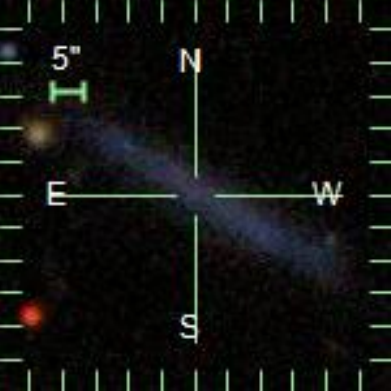}
\includegraphics[width=0.16\textwidth]{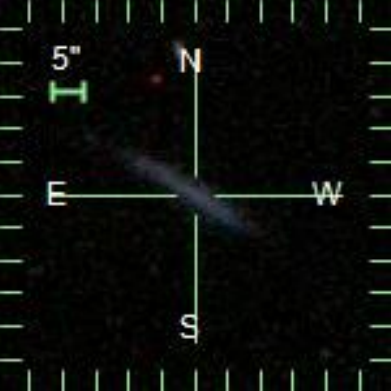}
\includegraphics[width=0.16\textwidth]{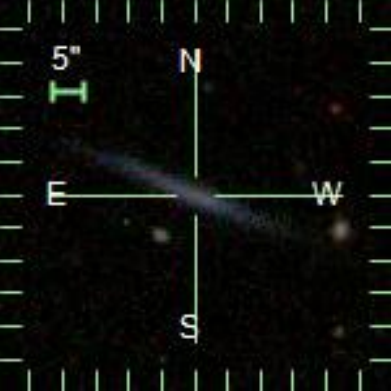}\\
\includegraphics[width=0.16\textwidth]{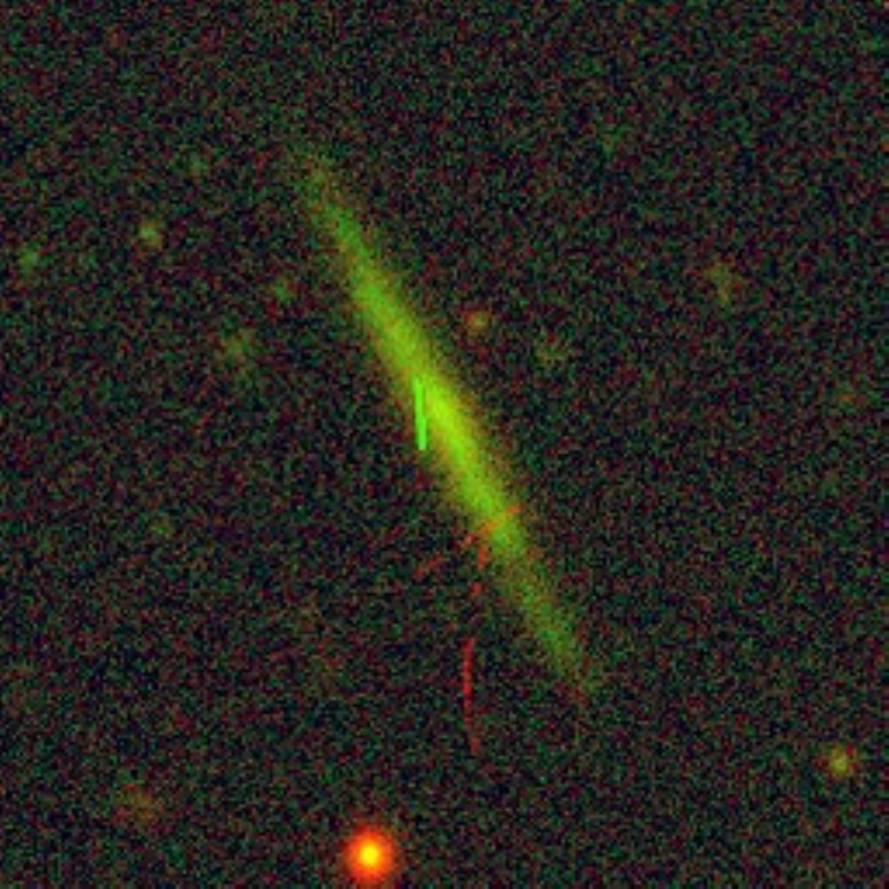}
\includegraphics[width=0.16\textwidth]{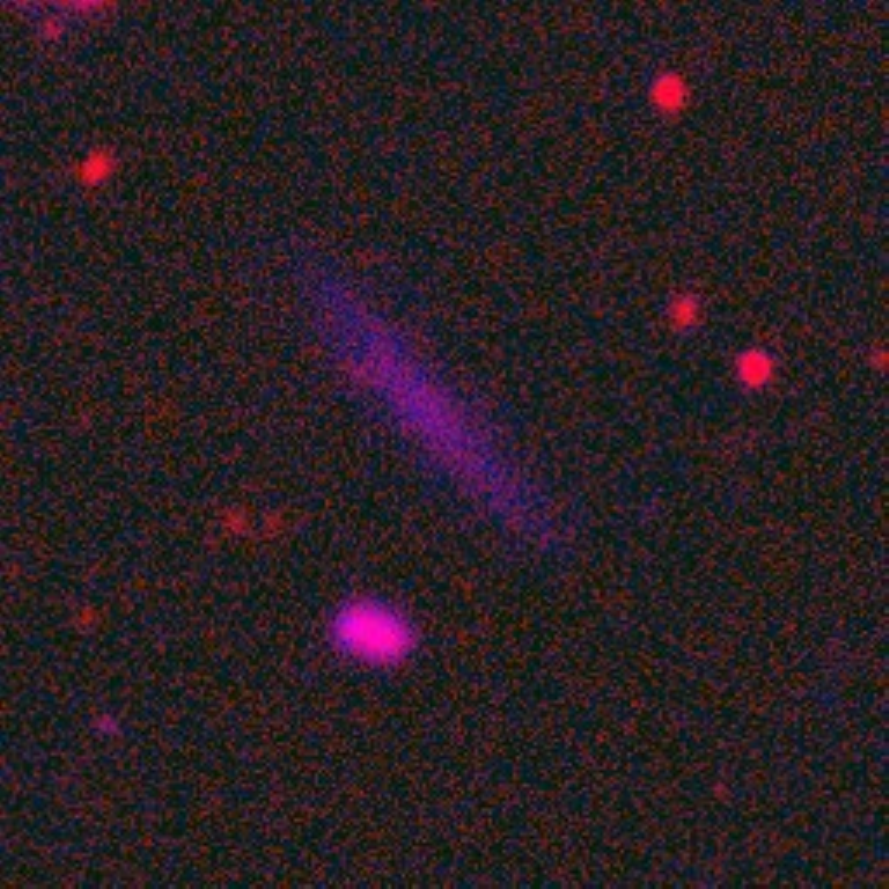}
\includegraphics[width=0.16\textwidth]{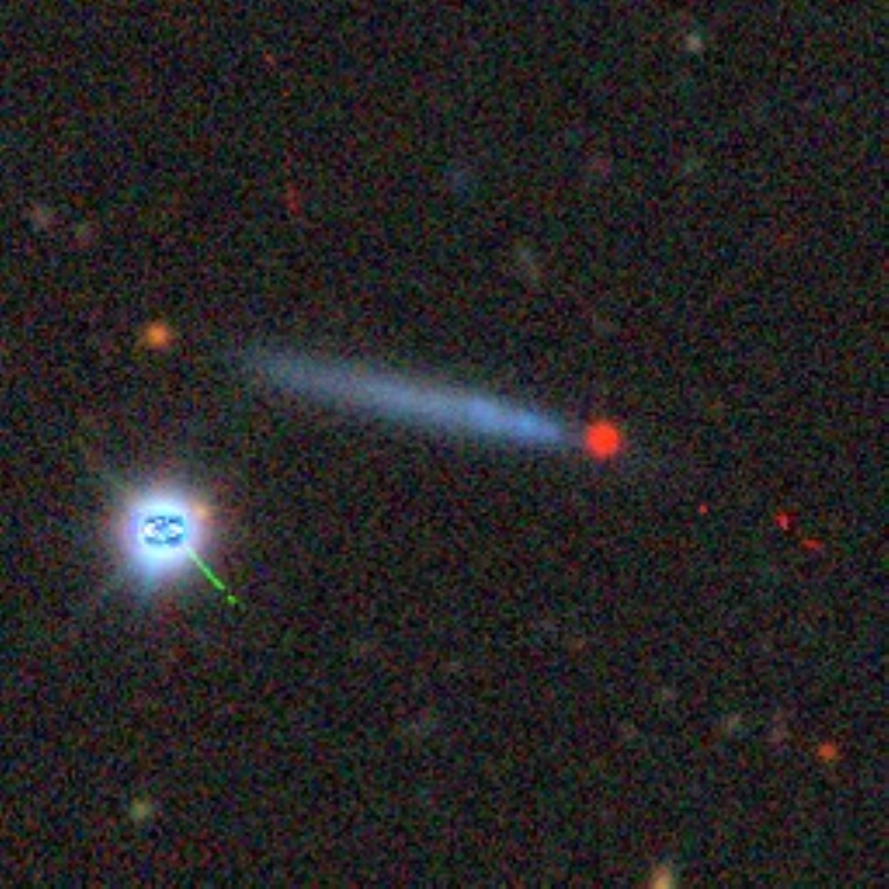}
\includegraphics[width=0.16\textwidth]{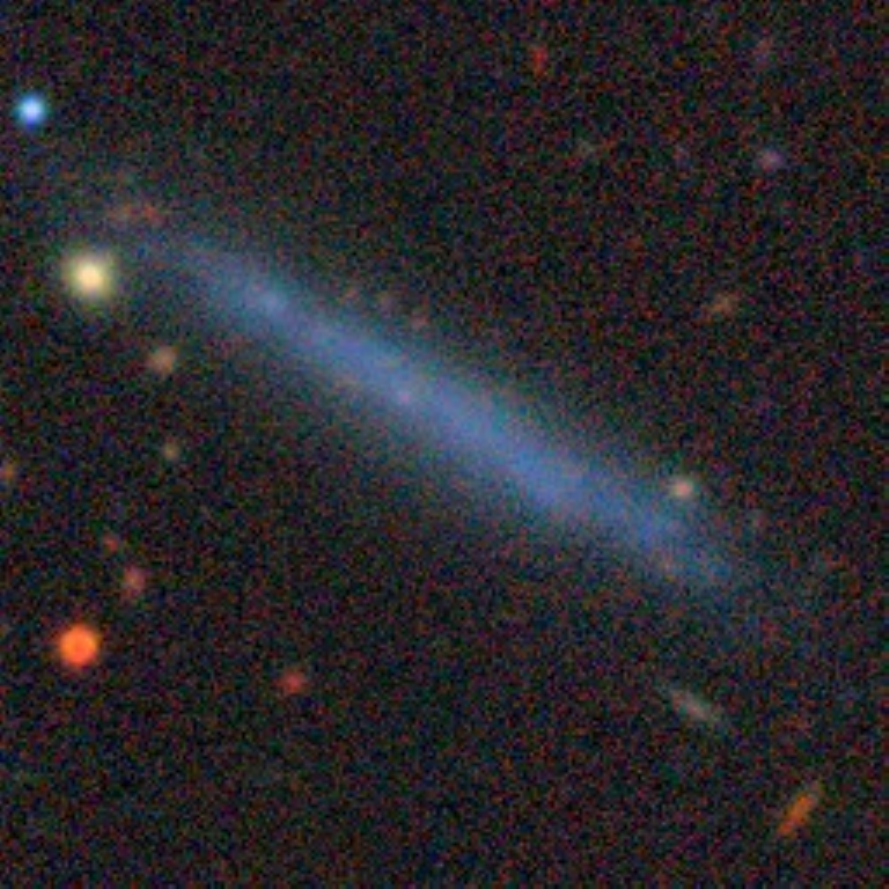}
\includegraphics[width=0.16\textwidth]{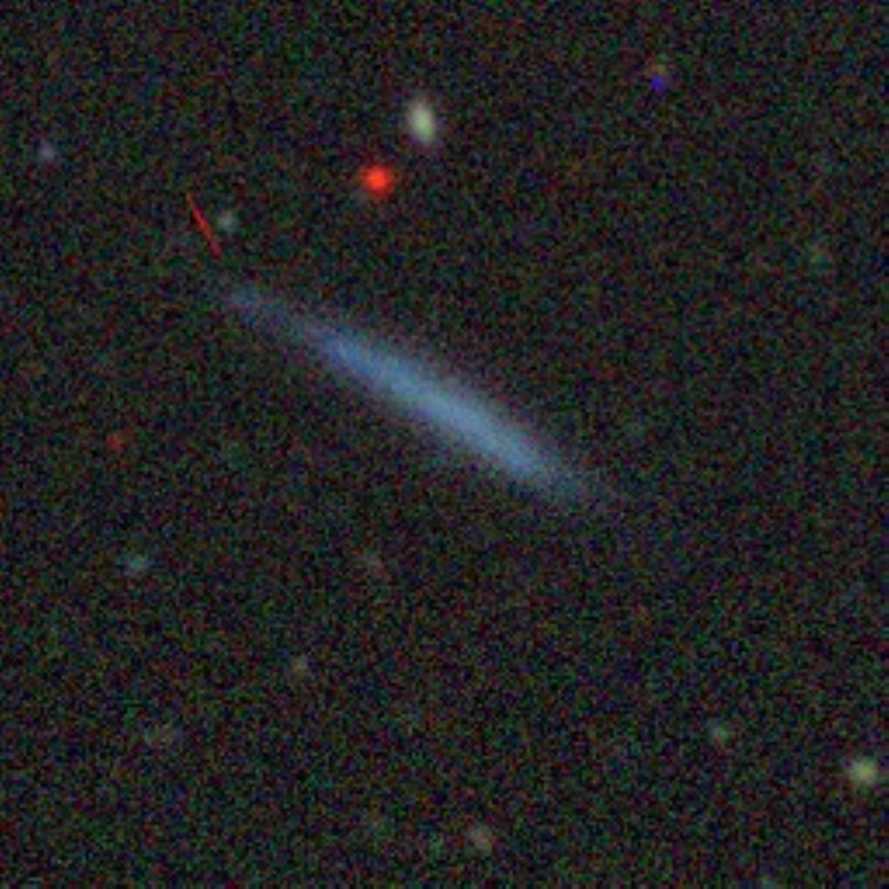}
\includegraphics[width=0.16\textwidth]{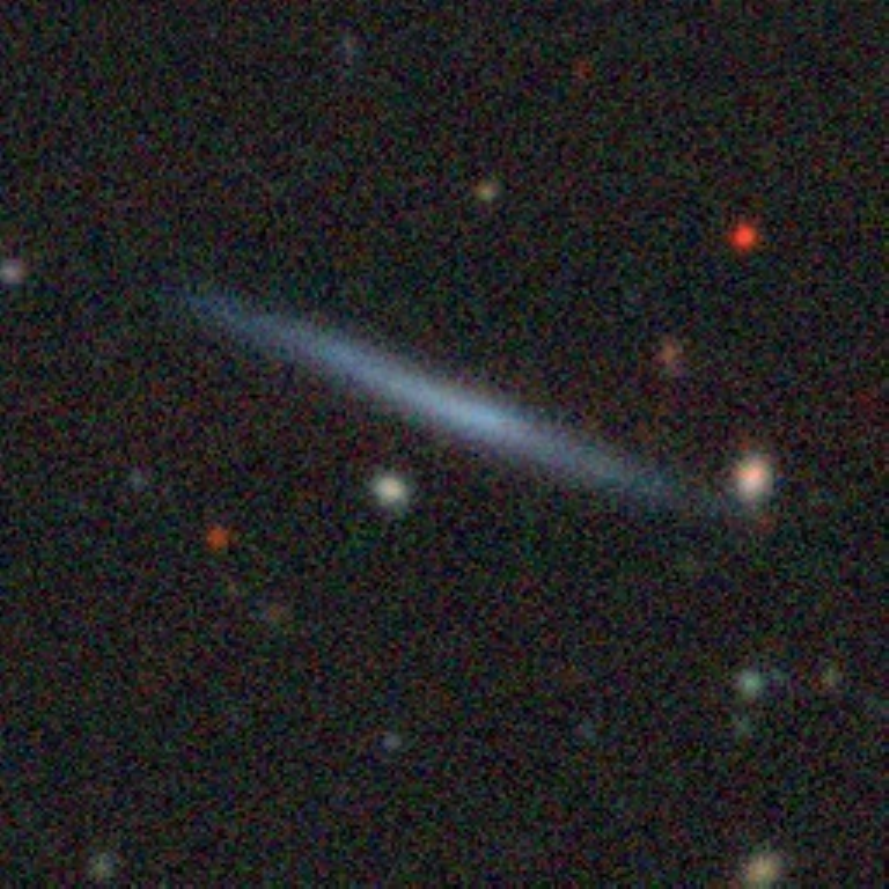}\\
\includegraphics[width=0.16\textwidth]{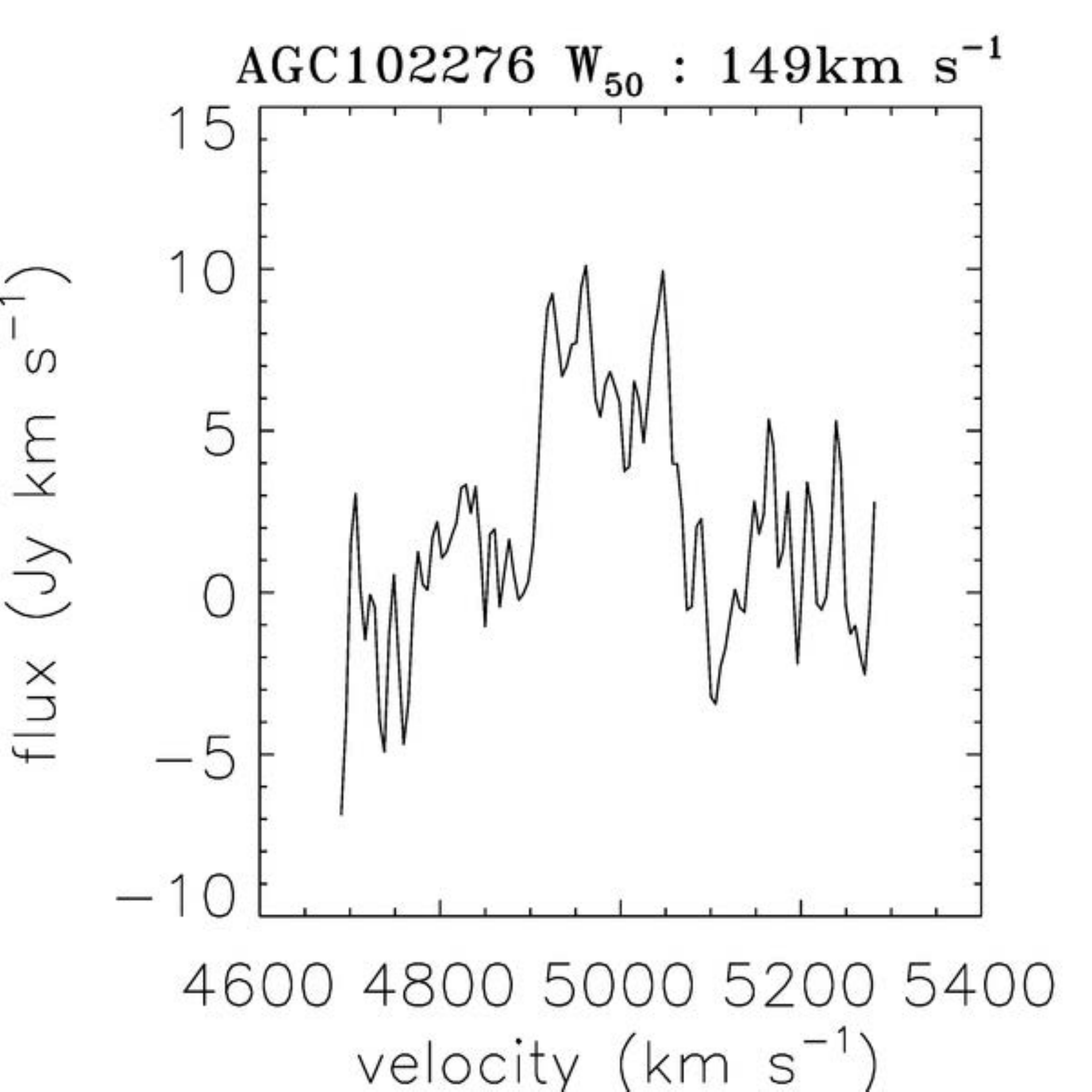}
\includegraphics[width=0.16\textwidth]{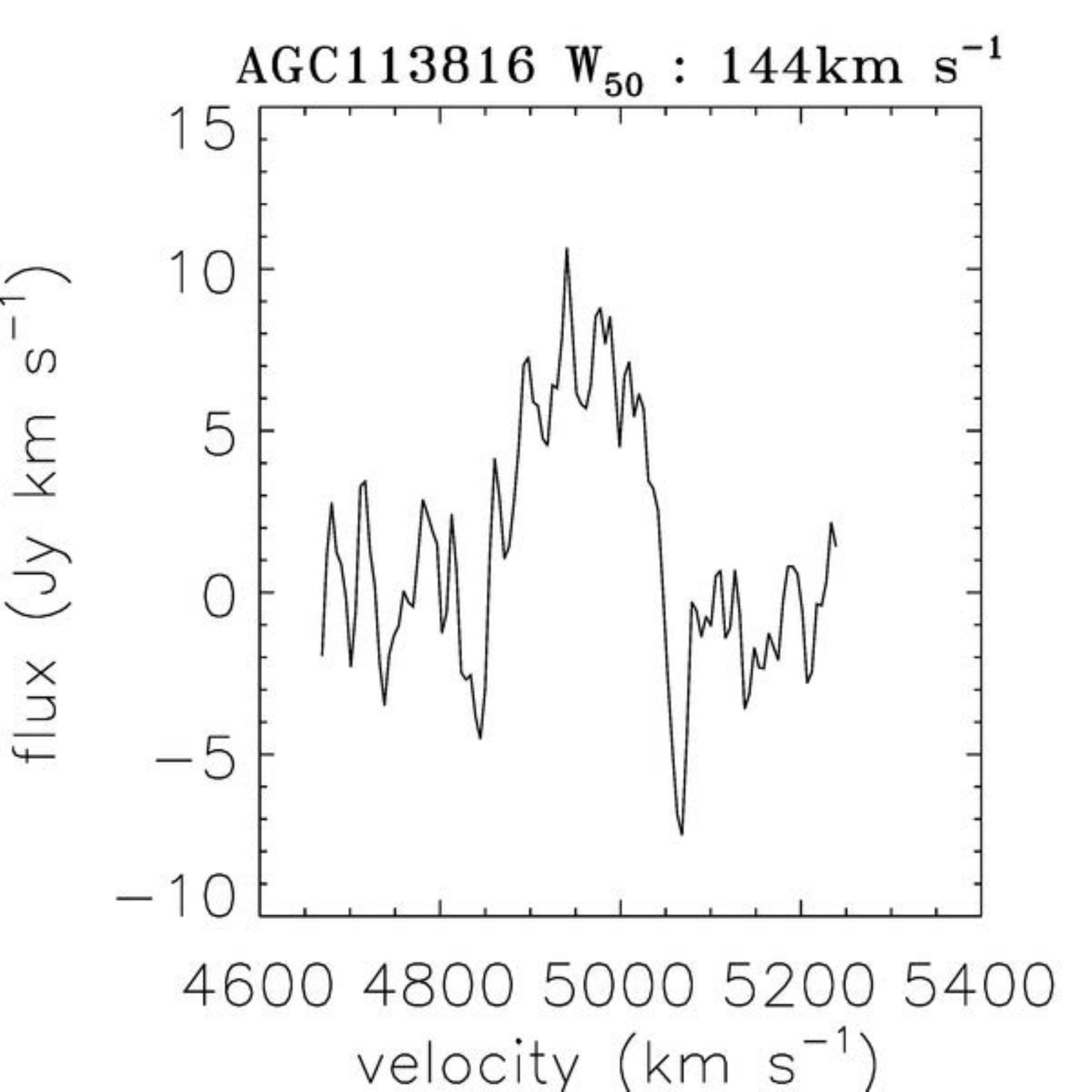}
\includegraphics[width=0.16\textwidth]{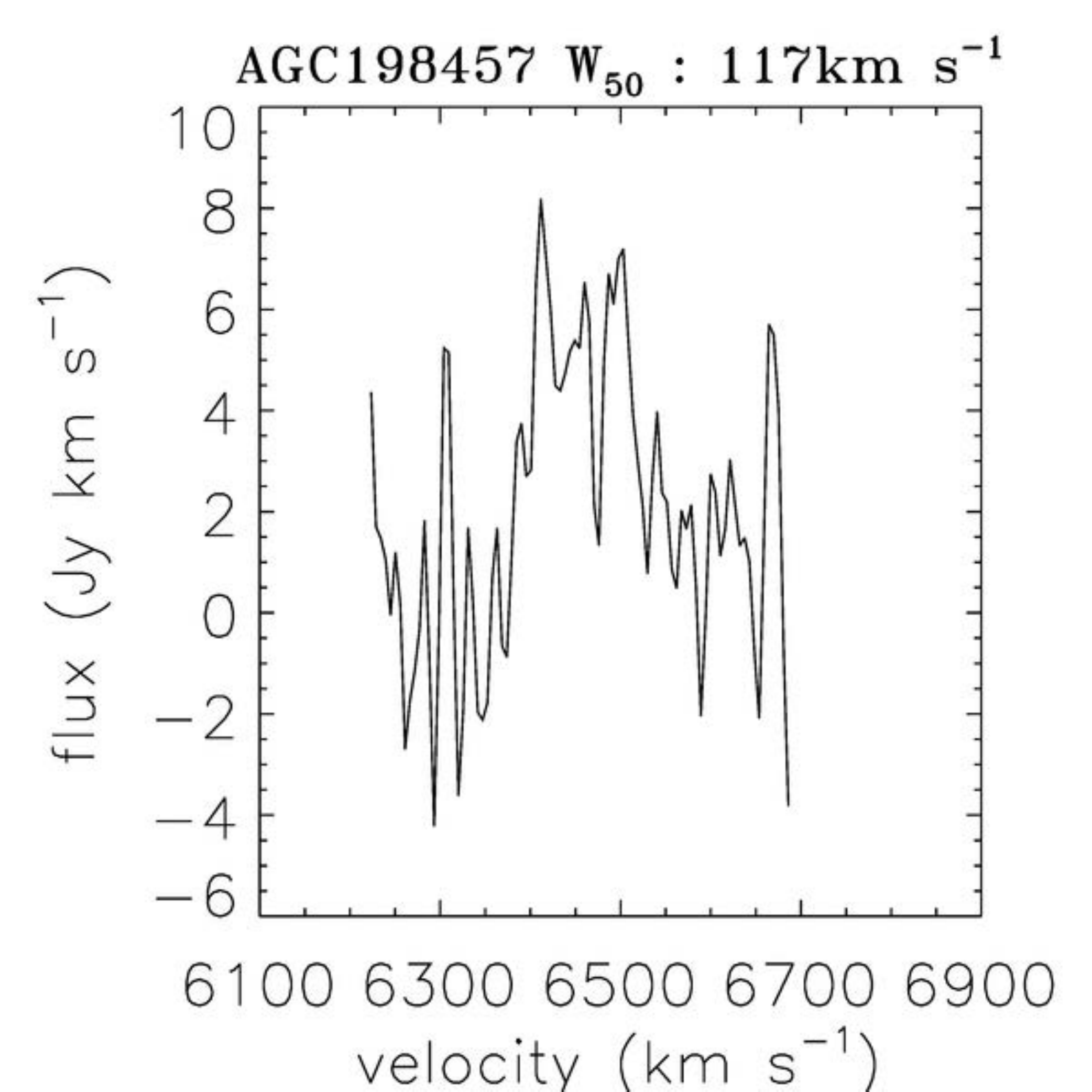}
\includegraphics[width=0.16\textwidth]{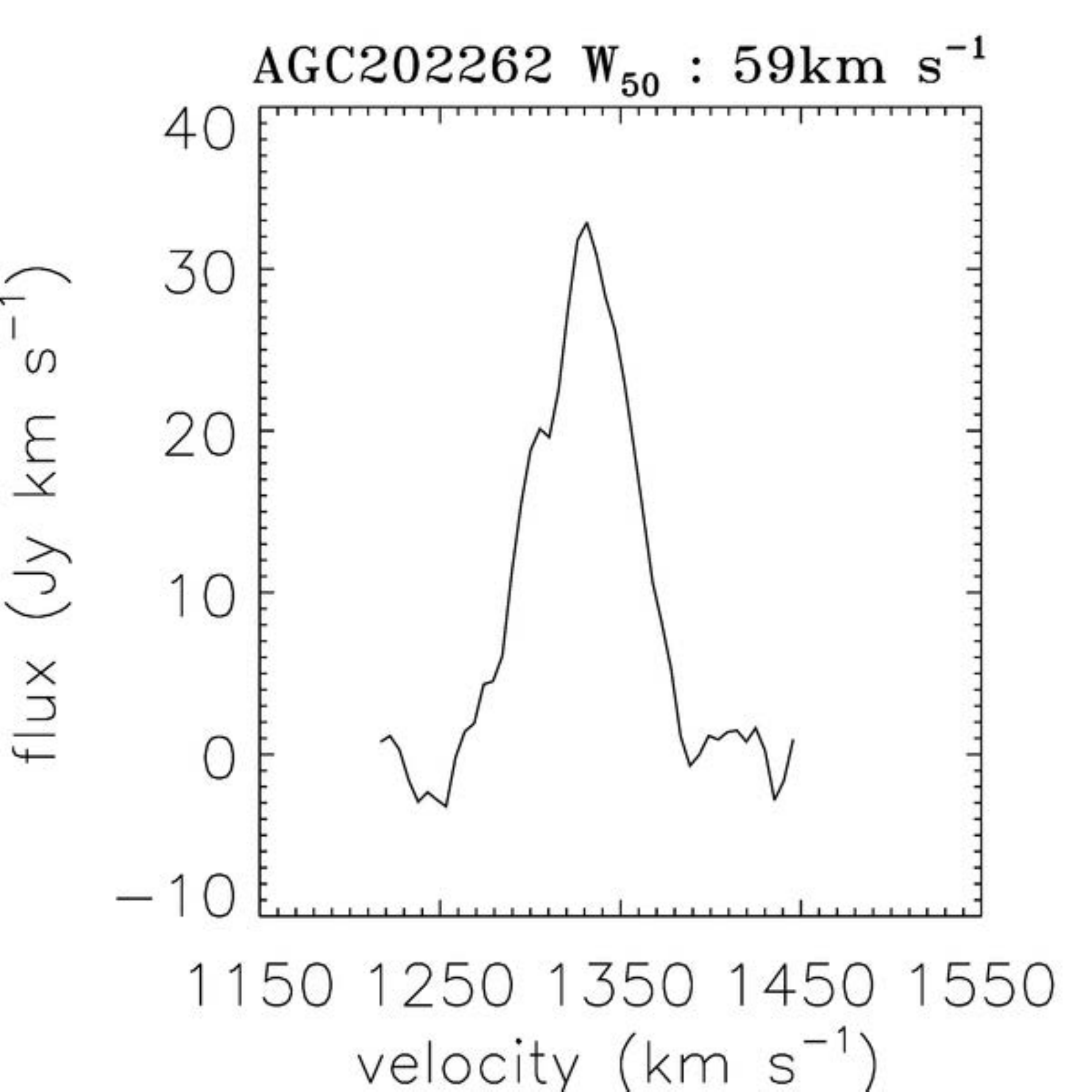}
\includegraphics[width=0.16\textwidth]{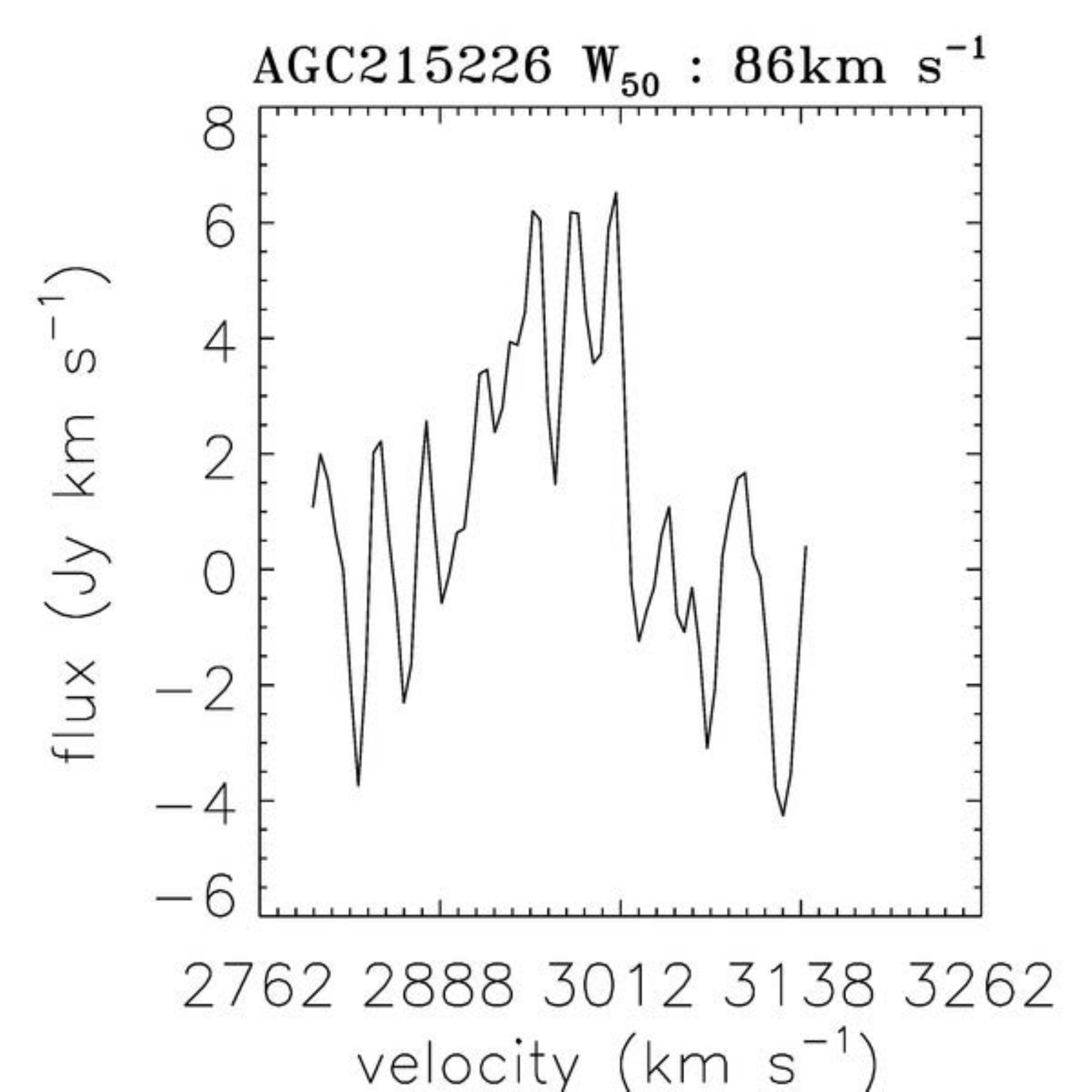}
\includegraphics[width=0.16\textwidth]{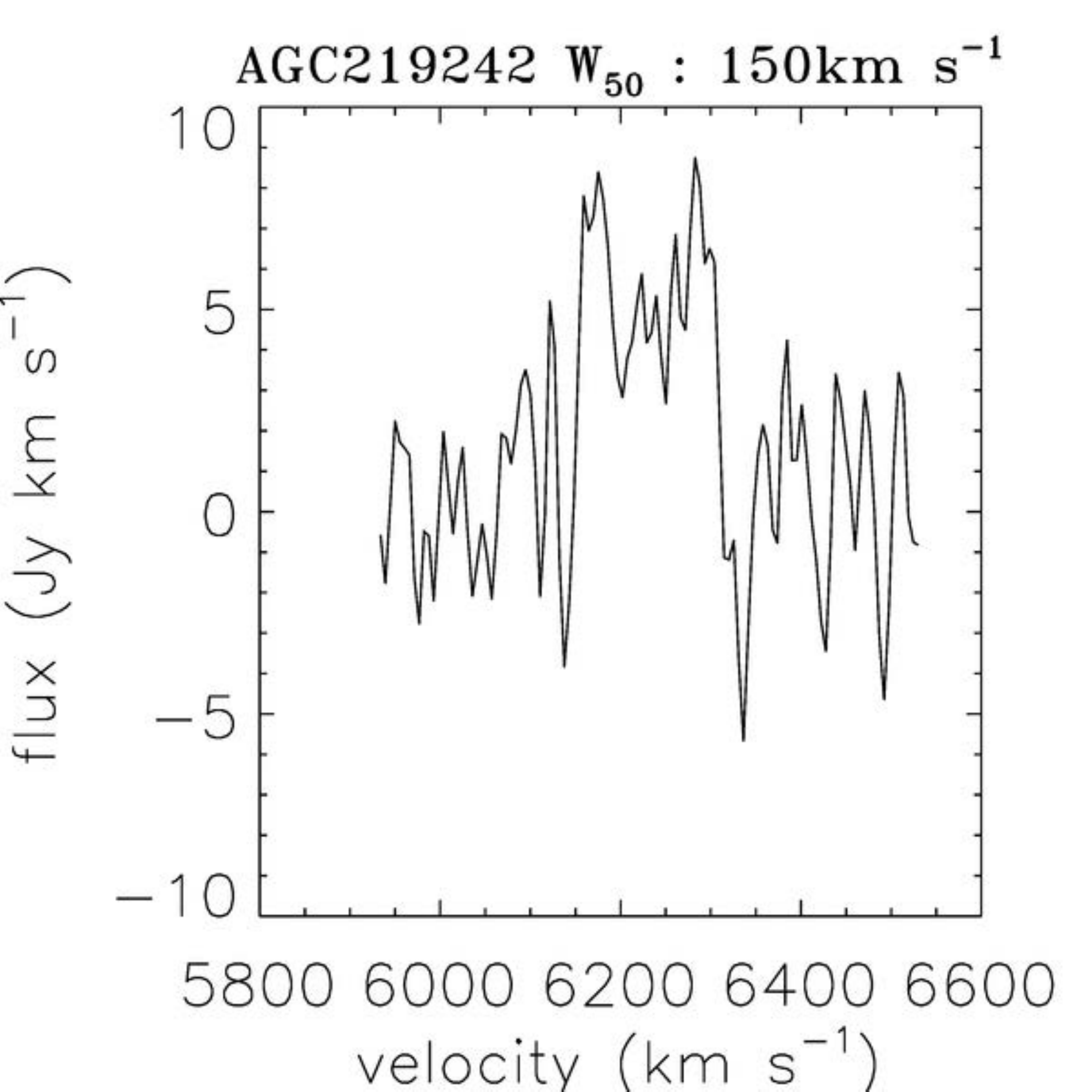}\\
\vspace{0.0cm}
{\bf \hspace{0.01in} AGC~223141 \hspace{0.26in} AGC~321194 \hspace{0.26in} AGC~729579 \hspace{0.26in} AGC~749223 \hspace{0.26in} AGC~749493}\\
\vspace{0.1cm}
\includegraphics[width=0.16\textwidth]{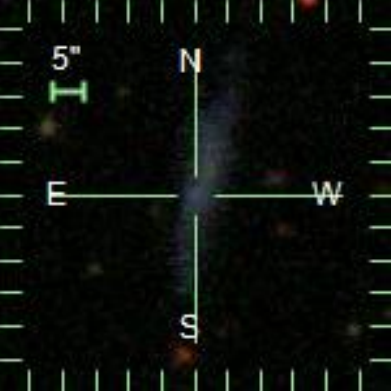}
\includegraphics[width=0.16\textwidth]{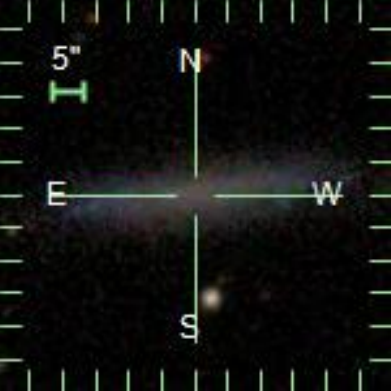}
\includegraphics[width=0.16\textwidth]{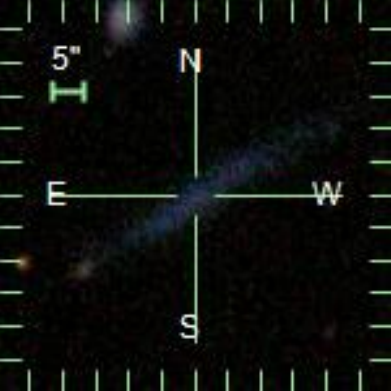}
\includegraphics[width=0.16\textwidth]{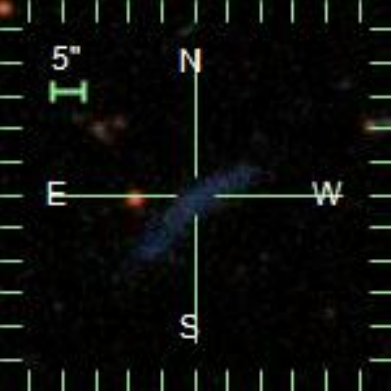}
\includegraphics[width=0.16\textwidth]{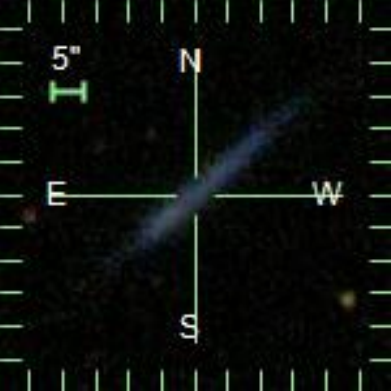}\\
\includegraphics[width=0.16\textwidth]{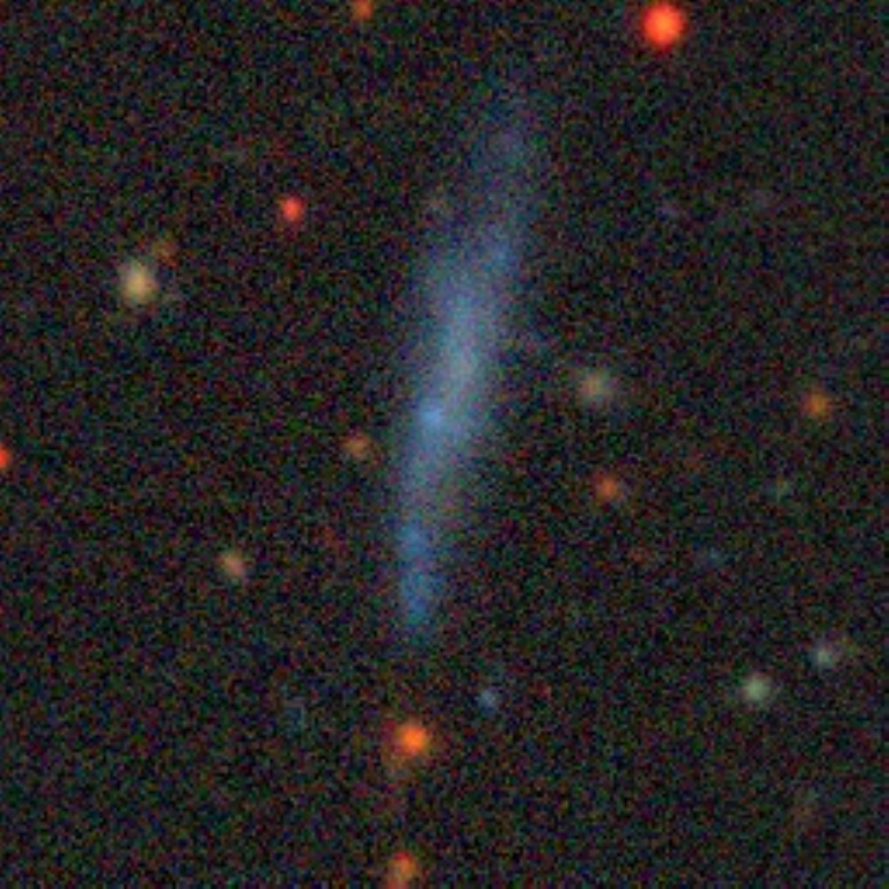}
\includegraphics[width=0.16\textwidth]{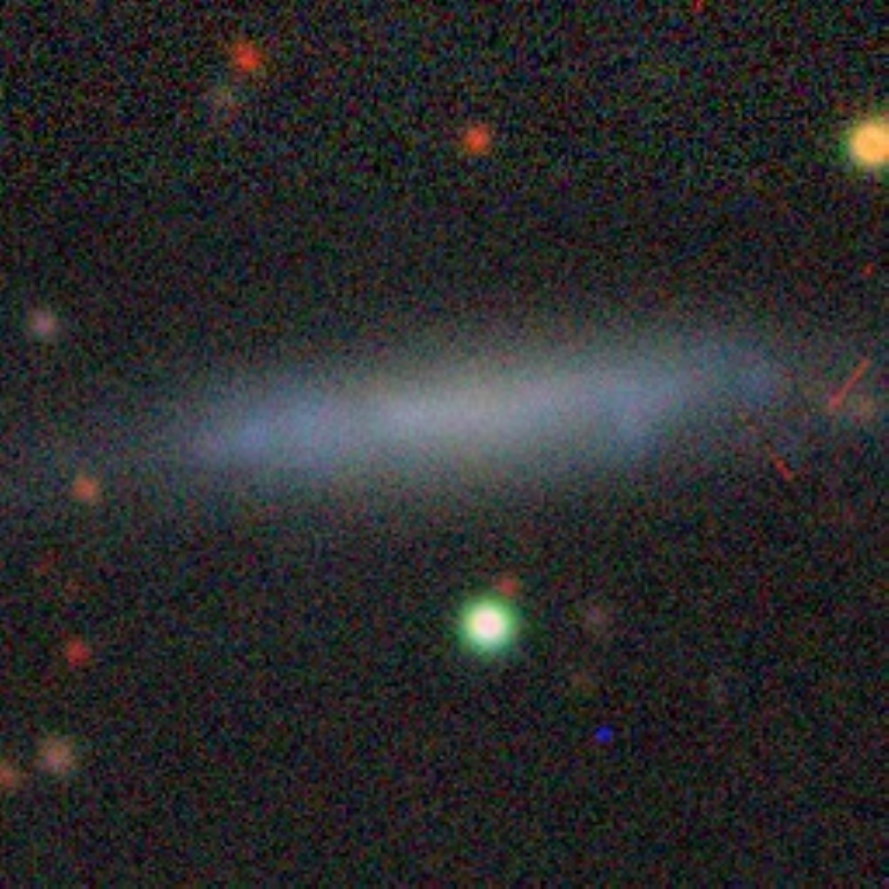}
\includegraphics[width=0.16\textwidth]{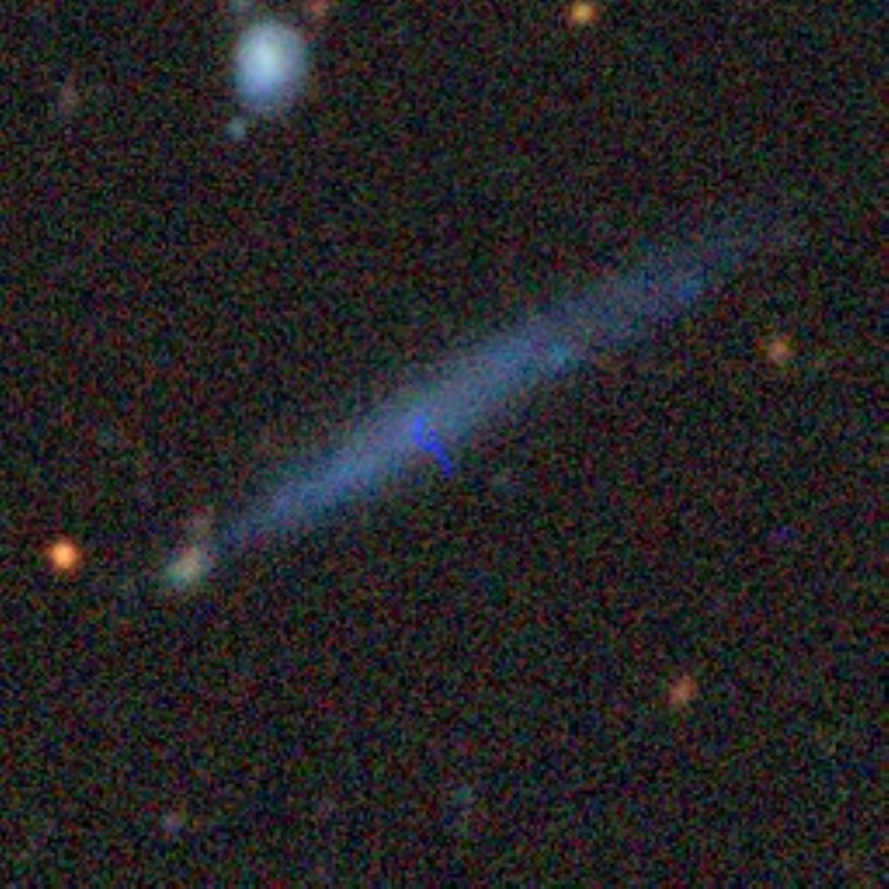}
\includegraphics[width=0.16\textwidth]{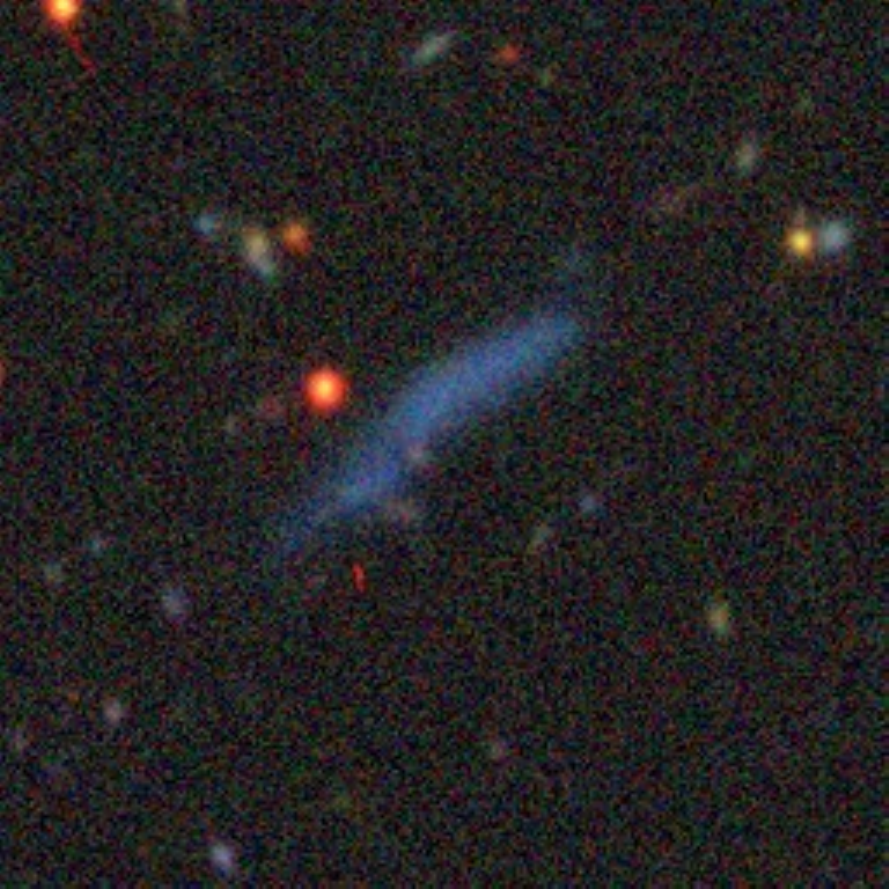}
\includegraphics[width=0.16\textwidth]{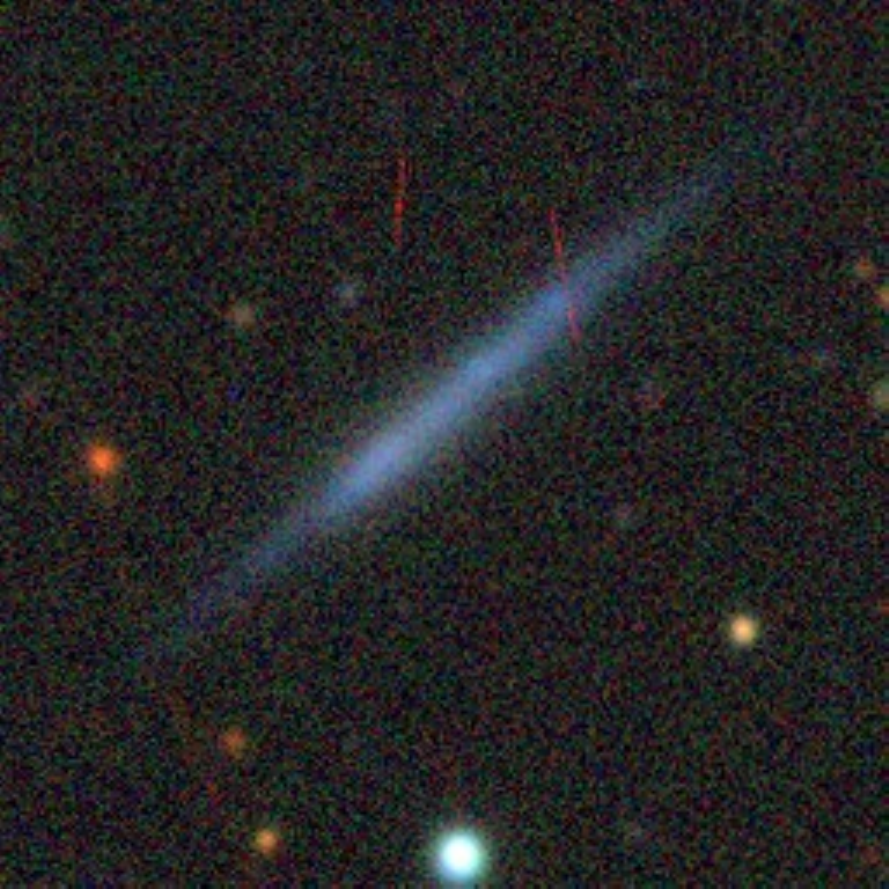}\\
\includegraphics[width=0.16\textwidth]{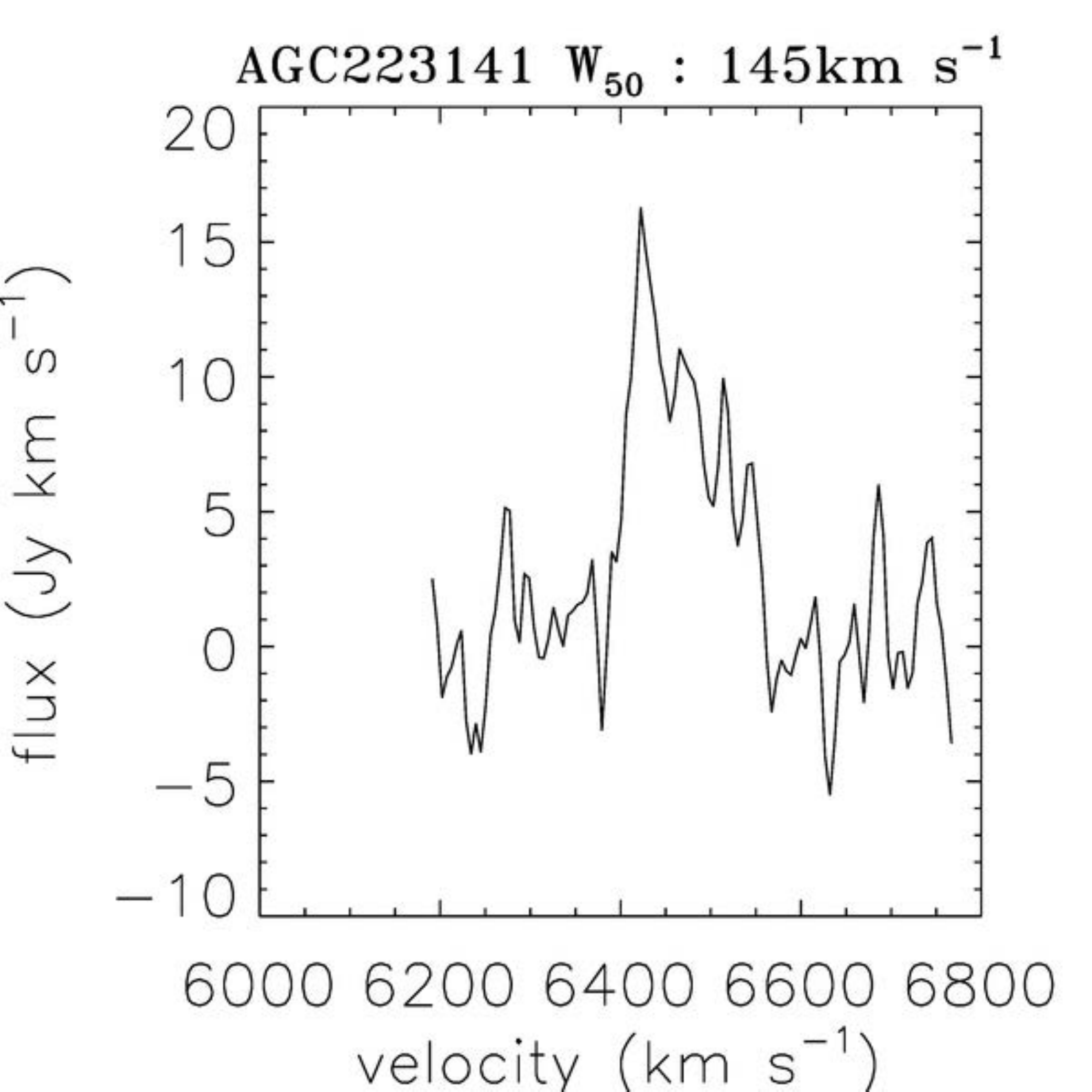}
\includegraphics[width=0.16\textwidth]{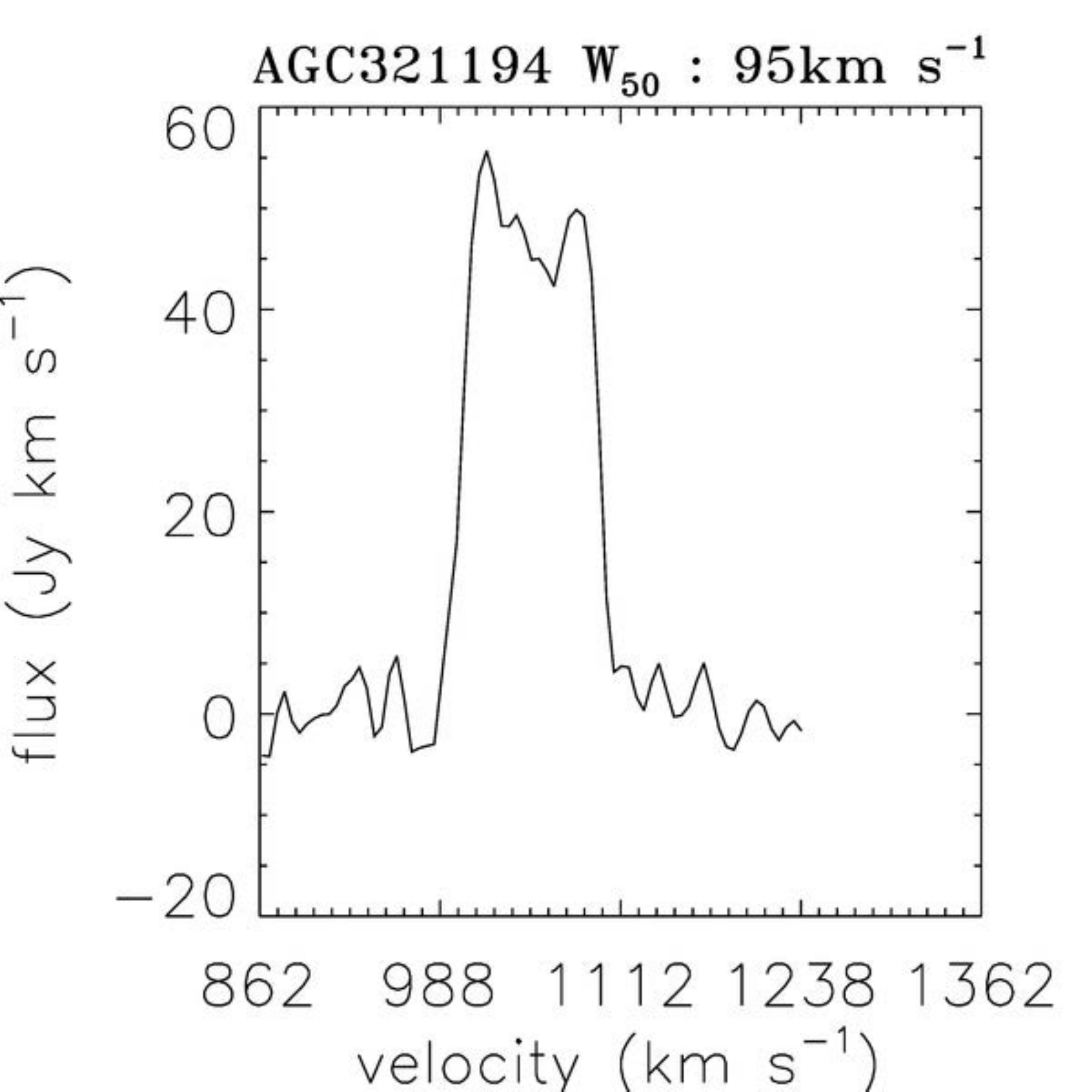}
\includegraphics[width=0.16\textwidth]{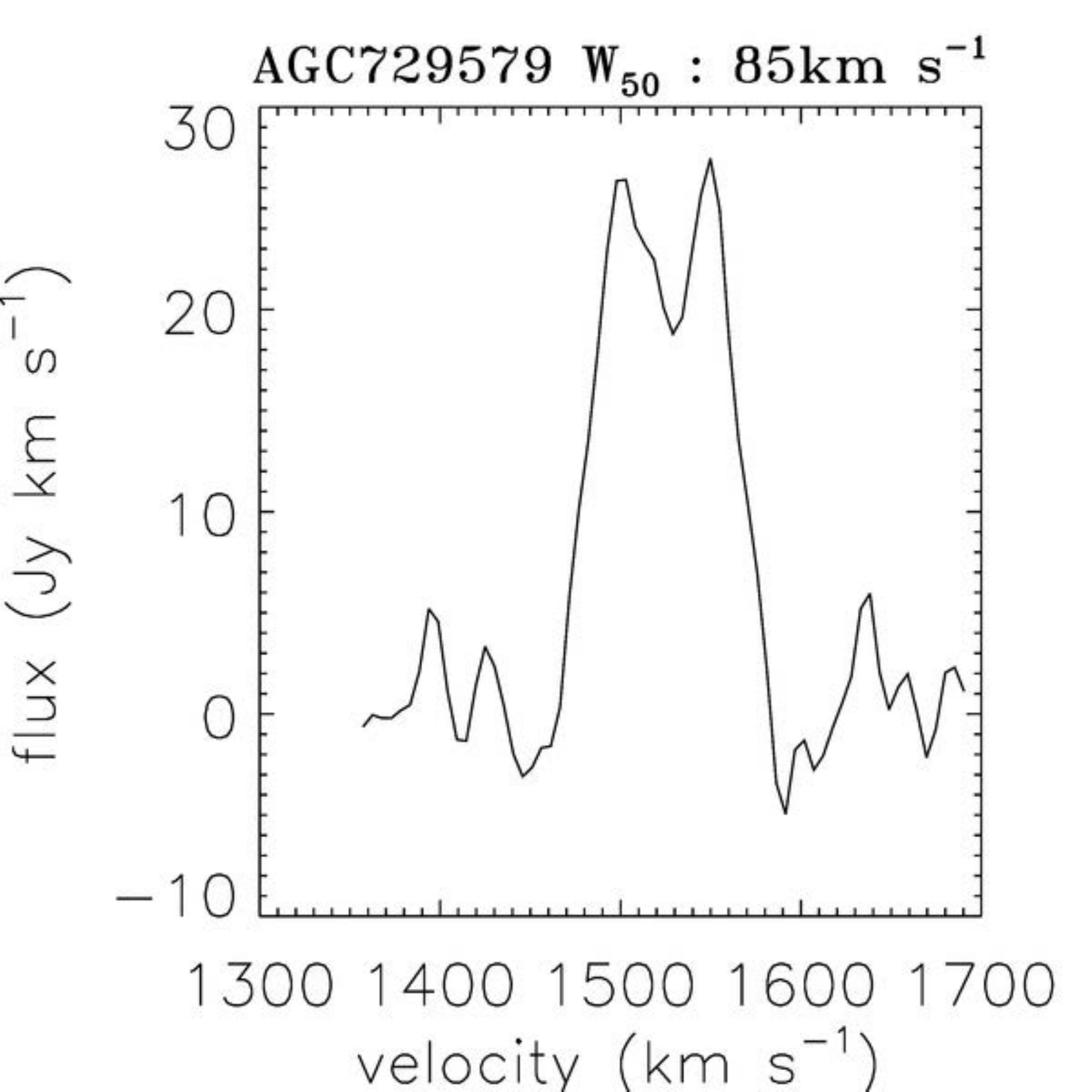}
\includegraphics[width=0.16\textwidth]{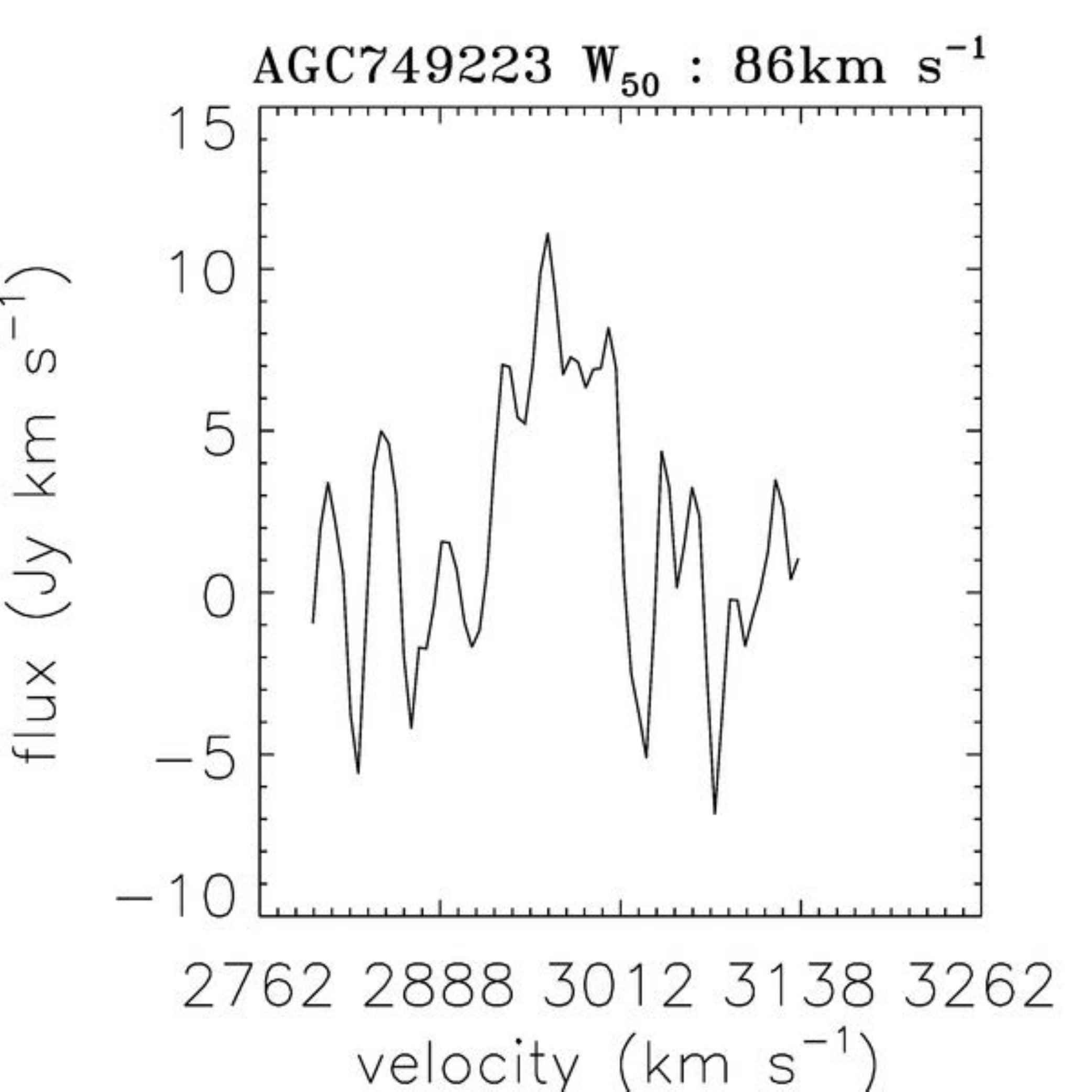}
\includegraphics[width=0.16\textwidth]{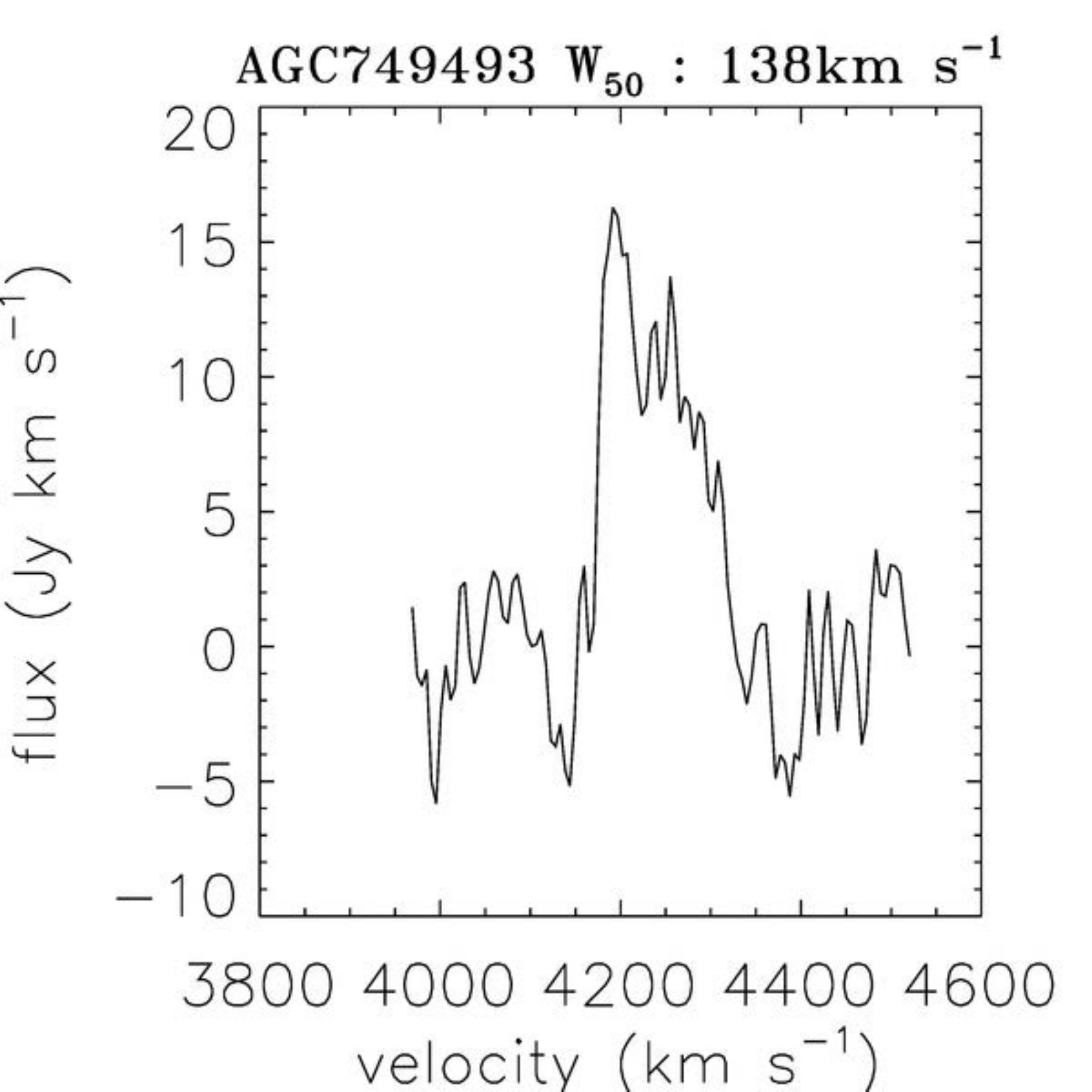}\\
\vspace{-0.2cm} \caption{These pictures show the SDSS DR7 images, DECaLS images and H{\sc{i}}-line spectra of the 11 edge-on HUDS candidates in the $\alpha .40$ catalog. SDSS DR7 images are created by combining the $g$-/$r$-/$i$-band images with blue, green and red colors respectively. The DECaLS images are targets in $g$-/$z$-bands and H{\sc{i}}-line spectra are achieved from the NASA Extragalactic Database. These optical images show edge-on disk-like morphologies and most of them exhibit the double-horned profiles of the H{\sc{i}}-line (except for AGC 202262). It may be concluded that most of our HUDS candidates are edge-on disk-like galaxies \citep{Bosma.1978PhDT.195B}.}
\label{fig:images} \vspace{0.4cm}
\end{figure*}

With correcting the central surface brightness $\mu_{0}$ and cosmological dimming effects \citep{Trachternach..2006A&A...458..341T, Zhong..2008MNRAS.391..986Z, Du_Wei_2015AJ..149..199}, we get the $\mu_{g,0,face}$ of these 1670 edge-on galaxies. Then, we select HUDS by using the criteria adopted by \citetalias{van.Dokkum..2015ApJ...798L..45V}: $r_{g,e} \geqslant 1.5$kpc and $\mu_{g,0} \geqslant 24$ mag arcsec$^{-2}$. The $r_{e}$ can be calculated by $r_{e} = 1.678\times r_{s}$ for a disk-like model fitting. There are only 11 galaxies that satisfy the criteria. Their SDSS DR7 and deeper DECaLS images, along with their H{\sc{i}}-line spectra, which are achieved from the NASA Extragalactic Database\footnote{\url{https://ned.ipac.caltech.edu/cgi-bin/NEDspectra?objname=&extend=multi&detail=0&preview=1&numpp=20&refcode=2011AJ....142..170H&bandpass=ANY&line=ANY}}, are shown in Figure \ref{fig:images} and their general parameters are listed in Table \ref{tab:params}. Absolute magnitudes of these galaxies are all fainter than $-17$ mag, and the color $g-r$ are bluer than $0.4$.\\

From Figure \ref{fig:images}, the globally integrated 21-cm emission lines of most of these HUDS candidates, except for AGC 202262, present double-horned profiles, which are a typical characteristic of disk galaxies \citep{Bosma.1978PhDT.195B}. Considering the symmetrical thin edge-on disk-like optical shapes, these characteristics of H{\sc{i}} velocity spectra and optical images support that most of these HUDS candidates are possible edge-on disk galaxies with rotating H{\sc{i}} gas.\\

\subsection{Dust extinction}
However, the existing dust in exponential profile systems would reduce the flux of edge-on galaxies. So generally, we could not transform the central surface brightness just by Equation \ref{eq:correct-mu} simplistically. However, it is difficult to know the exact extinction relation between face-on and edge-on galaxies, because of the unknown complex extinction inside the galaxies.\\

\begin{figure}
\plotone{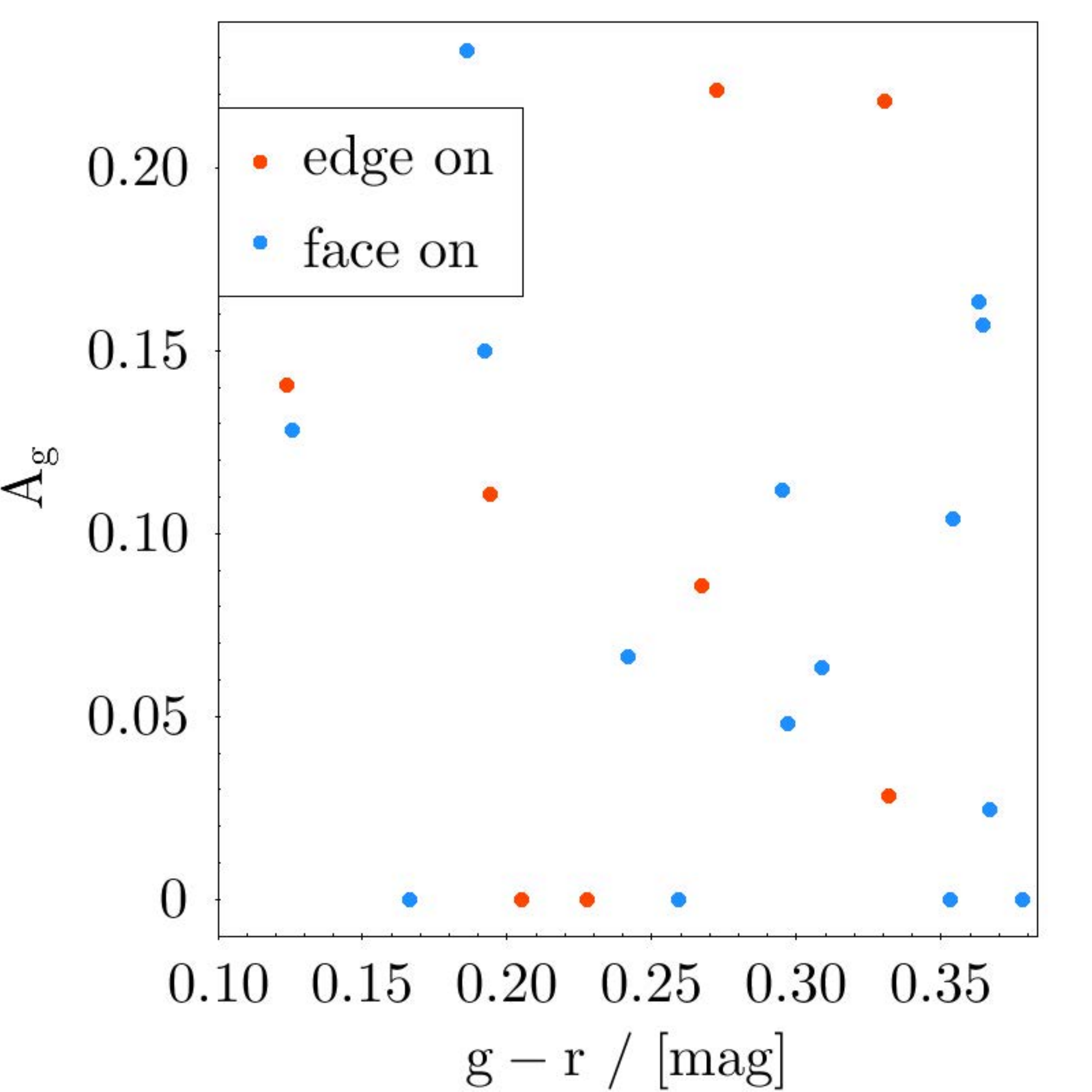}
\caption{This picture shows the $A_g$ of the edge-on (red dots) and face-on (blue dots) galaxies, which have good quality $H_{\alpha}$ and $H_{\beta}$ spectral information and are fainter than $-17$ absolute mag, with $g-r$ color bluer than $0.4$. There are not obvious differences between their distributions. This may indicate that the internal extinction of low luminosity edge-on galaxies does not differ very much from the low luminosity face-on galaxies.}
\label{fig:Ag}
\end{figure}

For estimating the internal extinction of our edge-on HUDS candidates, we have selected edge-on galaxies ($b/a<0.2$) and face-on galaxies ($b/a>0.8$) from the cross matches between the $\alpha .40$ catalog and the MPA-JHU DR7 catalog. These galaxies are further selected to be fainter than $-17$ mag and $g-r < 0.4$, which are the same as our edge-on HUDS candidates. Then, we removed those objects with poor spectral quality (having SNR of $H_{\beta}$ and $H_{\alpha}$ less than 5). By these selection criteria, eight edge-on galaxies and 15 face-on galaxies remain.\\

Then, we use the flux ratio of $H_{\alpha}$ and $H_{\beta}$, which are provided by the MPA-JHU DR7 catalog, to estimate the $g$-band internal extinction $A_g$ of each galaxy by using the widely-used Balmer decrement method. We apply the relationship between $H_{\alpha}$/$H_{\beta}$ and $E(B-V)$ \citep{Calzetti.2001PASP..113.1449C}, and then obtain the $A_g$:
\begin{equation}
E(B-V)=\frac{1.086/R_{V}}{\frac{A(H_{\beta})}{A_{V}}-\frac{A(H_{\alpha})}{A_{V}}}ln(\frac{H_{\alpha}/H_{\beta}}{H_{\alpha0}/H_{\beta0}}).
\end{equation}

For the galaxies which have $A_g < 0$, we assume they have $A_g = 0$. Finally, we plot the derived extinction $A_g$ of edge-on (red dots) and face-on (blue dots) galaxies in the Figure \ref{fig:Ag}. As shown in the figure, an obvious systematic offset between the extinction of the edge-on and face-on galaxies does not appear to exist.\\

After deriving the dust extinction value of edge-on and face-on galaxies, we further compare the dust extinction value between edge-on and face-on galaxies. Statistically, the mean values of $A_g$ are $0.10$ mag for edge-on galaxies and $0.08$ mag for face-on galaxies with the standard deviation ($\sigma$) of $0.08$ mag for edge-on galaxies and $0.07$ mag for face-on galaxies. The difference between the mean extinction values of edge-on and face-on galaxies is $0.02$ mag, which is much smaller than the standard deviations. This indicates that the internal extinction of edge-on galaxies may generally not differ much from that of face-on galaxies in terms of statistics. If we insist on correcting the dust extinction for edge-on galaxies, large uncertainties might be involved with the magnitudes of edge-on galaxies. Also as \cite{Giovanelli..1995AJ....110.1059G, Masters..2003AJ....126..158M, Maller..2009ApJ...691..394M, Masters.2010MNRAS.404..792M} and \cite{Devour..2016MNRAS.459.2054D} discuss, with the decrease in luminosity of galaxies, the attenuation parameter $\gamma$ declines and the variety of relative extinction in $g$-band flattens. These suggest that low luminosity galaxies tend to be low extinction. So, we decide not to correct the internal extinction of our edge-on galaxies. \\

\section{Properties of HUDS candidates}
\label{sec:results}

\begin{table*}[ht!]
\centering
\caption{Parameters of the edge-on HUDS candidates}
\label{tab:params}
\setlength{\leftskip}{-30pt}
\resizebox{\textwidth}{!}{
\centering
\begin{tabular}{ccccccccccccccc}
\hline
\hline
 AGCNr
 & RA.
 & Dec.
 & $M_{g}$ \footnote{\scriptsize Absolution magnitude calculated from photometry by using SExtractor and Distance provided from $\alpha .40$ catalog.}
 & $g-r$
 & $r_{g,e}$
 & $\mu_{g,0,edge}$\footnote{\scriptsize Observational edge-on perspective central surface brightness obtained from GALFIT fitting with edge-on disk model with correction of cosmological dimming effects.}
 & $\mu_{g,0,face}$\footnote{\scriptsize Face-on perspective central surface brightness corrected from observed edge-on values by a factor of $2.5lg(h_s/r_s)$.}
 & $cz$
 & Dist\footnote{\scriptsize Dist, $W_{50}$ and H{\sc{i}} mass are achieved directly from the $\alpha. 40$ catalog.}
 & $W_{50}$
 & $lg(M_{H{\sc{I}}}/M_\odot)$
 & $lg(M_\star/M_\odot)$\footnote{\scriptsize Stellar mass calculated by using the $g$-band parameters in the work of \citet{Zibetti..2009MNRAS.400.1181Z} and $g-r$ color.}
 & $M_{H{\sc{I}}}$/$M_\star$
 & $\epsilon$\footnote{\scriptsize Ellipticity of HUDS candidates ($1-\frac{b}{a}$).}\\

 & (J2000) & (J2000)
 & (mag) & (mag)
 & (kpc)
 & (mag arcsec$^{-2}$) & (mag arcsec$^{-2}$)
 & (km s$^{-1}$)
 & (Mpc)
 & (km s$^{-1}$)
 & & & &\\
\hline
102276 & 00:49:52 & +25:56:39 & $-16.51\pm 0.07$ & $0.35\pm 0.02$ & $4.05\pm 0.14$ & $22.48\pm 0.01$ & $24.01\pm 0.02$ & 4984 & $69.6\pm 2.2$ & $149\pm 4 $ & $9.10\pm 0.06$ & $8.35\pm 0.05$ & $5.68 \pm  1.05$ & $0.91\pm 0.01$\\
113816 & 01:18:47 & +24:27:14 & $-15.46\pm 0.08$ & $0.40\pm 0.05$ & $3.25\pm 0.15$ & $23.56\pm 0.02$ & $24.64\pm 0.06$ & 4956 & $68.7\pm 2.3$ & $144\pm 13$ & $9.06\pm 0.06$ & $8.02\pm 0.11$ & $10.87\pm  3.19$ & $0.80\pm 0.03$\\
198457 & 09:53:00 & +07:15:22 & $-16.79\pm 0.06$ & $0.16\pm 0.04$ & $4.35\pm 0.13$ & $22.54\pm 0.02$ & $24.03\pm 0.04$ & 6457 & $97.0\pm 2.3$ & $117\pm 12$ & $9.20\pm 0.06$ & $8.06\pm 0.08$ & $13.81\pm  3.20$ & $0.87\pm 0.01$\\
202262 & 10:37:29 & +12:23:46 & $-14.58\pm 0.24$ & $0.22\pm 0.03$ & $1.99\pm 0.22$ & $22.95\pm 0.01$ & $24.31\pm 0.02$ & 1330 & $22.0\pm 2.4$ & $59 \pm 4 $ & $8.35\pm 0.10$ & $7.29\pm 0.11$ & $11.40\pm  3.90$ & $0.85\pm 0.01$\\
215226 & 11:43:32 & +15:02:33 & $-14.79\pm 0.12$ & $0.33\pm 0.05$ & $1.92\pm 0.11$ & $22.69\pm 0.02$ & $24.14\pm 0.04$ & 2970 & $45.0\pm 2.3$ & $86 \pm 12$ & $8.33\pm 0.07$ & $7.62\pm 0.11$ & $5.14 \pm  1.51$ & $0.86\pm 0.01$\\
219242 & 11:29:39 & +07:47:36 & $-16.77\pm 0.06$ & $0.30\pm 0.04$ & $4.28\pm 0.14$ & $22.33\pm 0.02$ & $24.04\pm 0.04$ & 6230 & $94.1\pm 2.4$ & $150\pm 8 $ & $9.26\pm 0.06$ & $8.34\pm 0.08$ & $8.39 \pm  1.90$ & $0.90\pm 0.01$\\
223141 & 12:41:12 & +10:55:58 & $-16.90\pm 0.06$ & $0.28\pm 0.03$ & $5.67\pm 0.17$ & $22.95\pm 0.01$ & $24.06\pm 0.03$ & 6479 & $97.2\pm 2.3$ & $145\pm 1 $ & $9.49\pm 0.05$ & $8.34\pm 0.07$ & $13.99\pm  2.82$ & $0.80\pm 0.01$\\
321194 & 22:57:04 & +25:43:30 & $-14.98\pm 0.29$ & $0.36\pm 0.02$ & $1.54\pm 0.21$ & $22.78\pm 0.01$ & $24.06\pm 0.02$ & 1050 & $16.5\pm 2.2$ & $95 \pm 2 $ & $8.48\pm 0.12$ & $7.76\pm 0.12$ & $5.27 \pm  2.07$ & $0.86\pm 0.01$\\
729579 & 11:24:17 & +25:42:38 & $-13.94\pm 0.21$ & $0.21\pm 0.04$ & $1.86\pm 0.18$ & $23.23\pm 0.02$ & $24.74\pm 0.04$ & 1523 & $25.4\pm 2.4$ & $85 \pm 3 $ & $8.50\pm 0.09$ & $7.02\pm 0.12$ & $30.02\pm 10.24$ & $0.87\pm 0.01$\\
749223 & 11:54:45 & +26:00:18 & $-14.80\pm 0.11$ & $0.02\pm 0.05$ & $2.18\pm 0.12$ & $23.16\pm 0.02$ & $24.25\pm 0.04$ & 2968 & $45.0\pm 2.3$ & $86 \pm 7 $ & $8.49\pm 0.07$ & $6.98\pm 0.11$ & $32.67\pm  9.88$ & $0.79\pm 0.01$\\
749493 & 15:05:05 & +24:02:04 & $-16.23\pm 0.08$ & $0.28\pm 0.03$ & $4.15\pm 0.16$ & $22.58\pm 0.01$ & $24.27\pm 0.03$ & 4245 & $64.2\pm 2.2$ & $138\pm 7 $ & $9.16\pm 0.06$ & $8.07\pm 0.07$ & $12.19\pm  2.51$ & $0.89\pm 0.01$\\
\hline
\hline
\end{tabular}}
\vspace{-0.2cm}
\begin{flushleft}
\textbf{\footnotesize Notes}.\\
{\scriptsize Errors of $M_g$ are RMS errors obtained by SExtractor, errors of $\mu_{g,0,edge}$ and $r_e$ are obtained by GALFIT, errors of H{\sc{I}} mass, Distance and $w_{50}$ are provided from $\alpha .40$ catalog, and the other errors are calculated by error propagation functions.}
\end{flushleft}
\vspace{-0.5cm}
\end{table*}

\begin{figure*}
\centering
\includegraphics[width=0.24\textwidth]{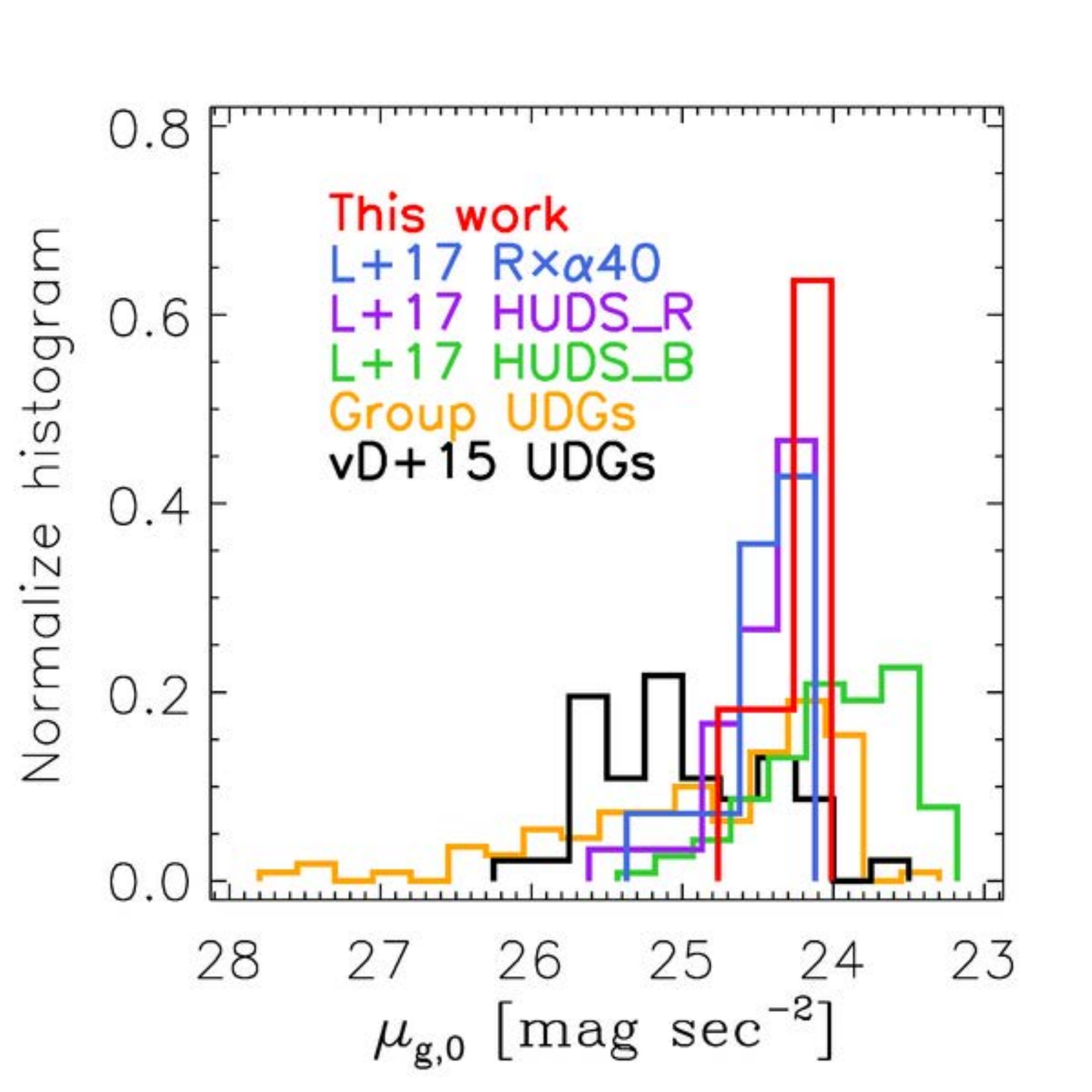}
\includegraphics[width=0.24\textwidth]{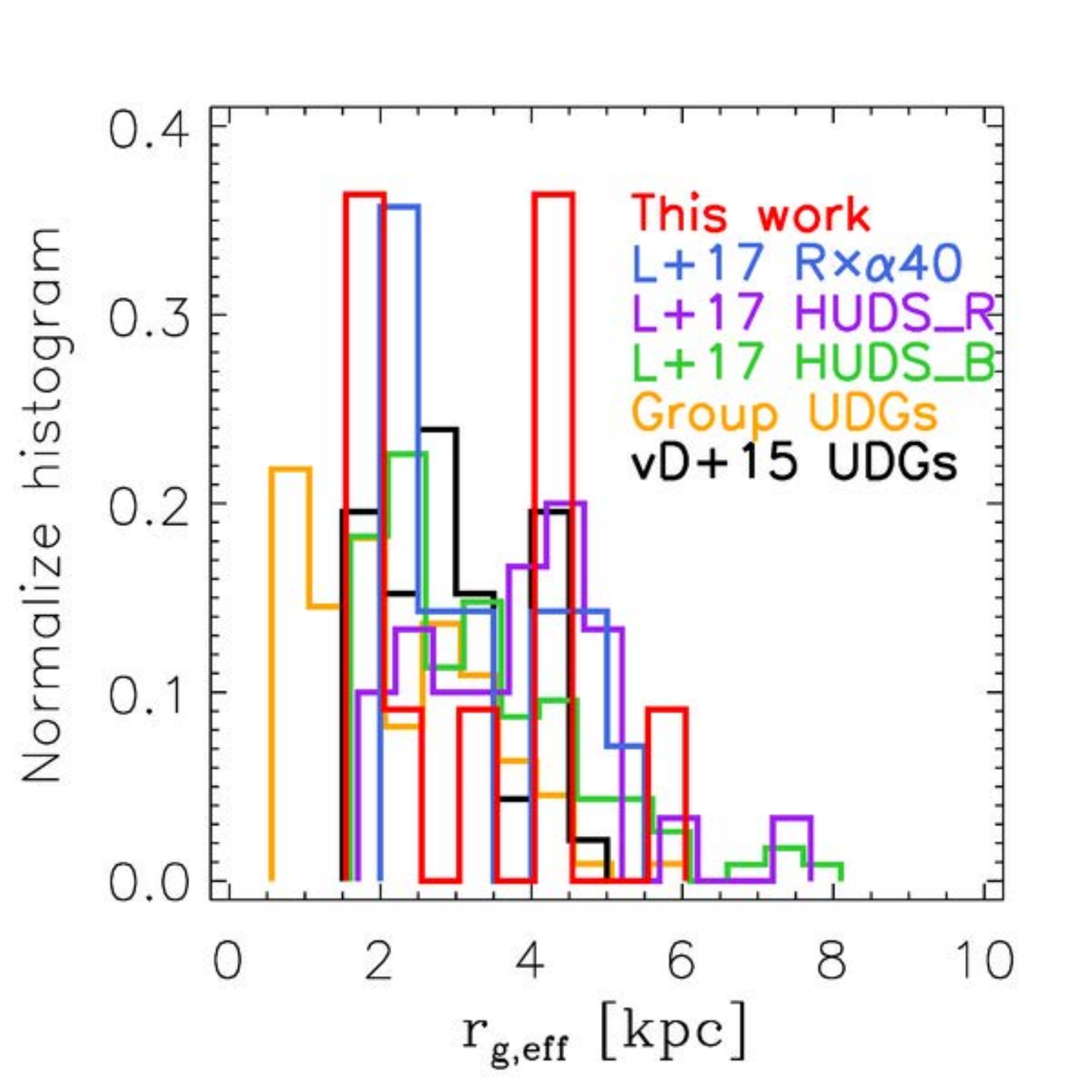}
\includegraphics[width=0.24\textwidth]{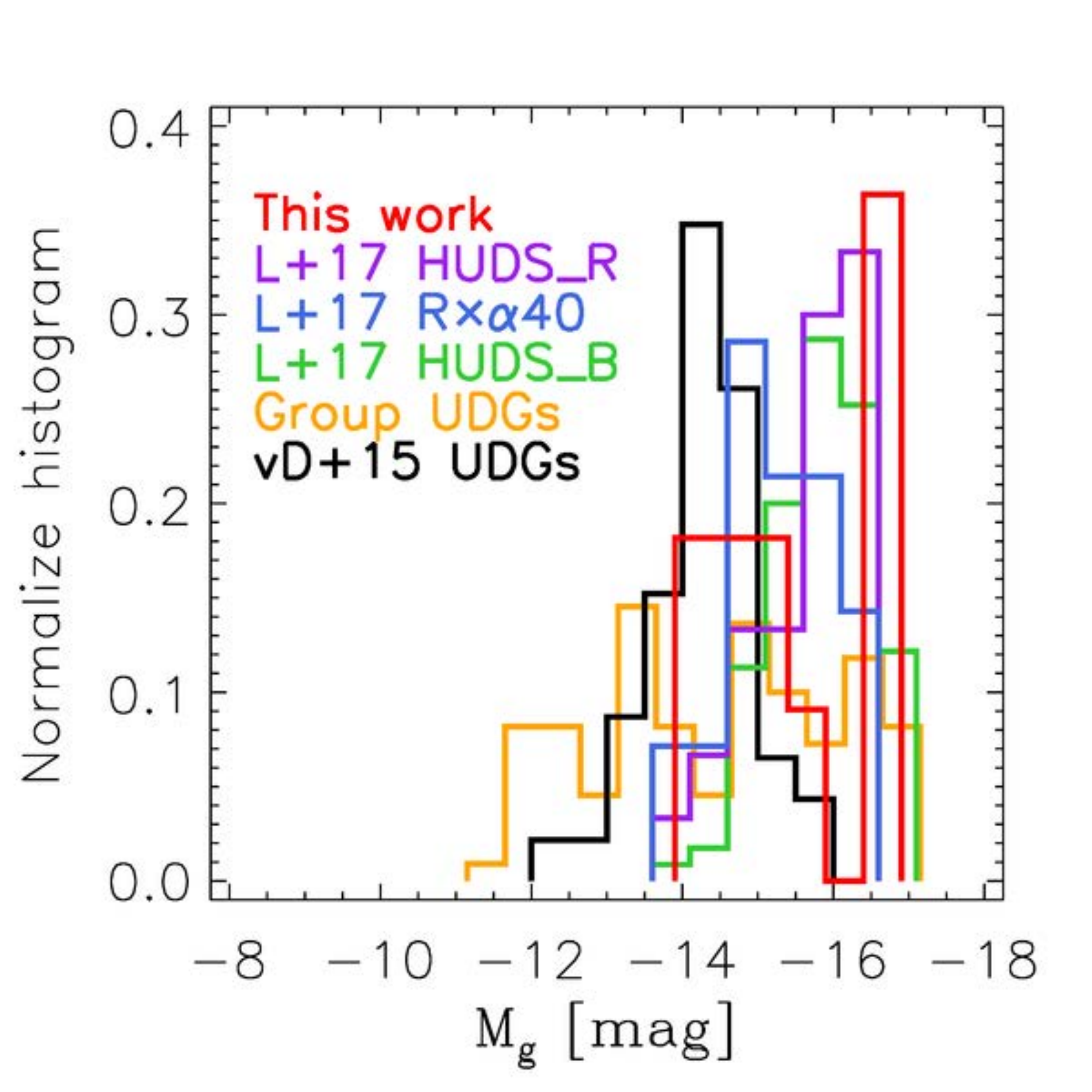}
\includegraphics[width=0.24\textwidth]{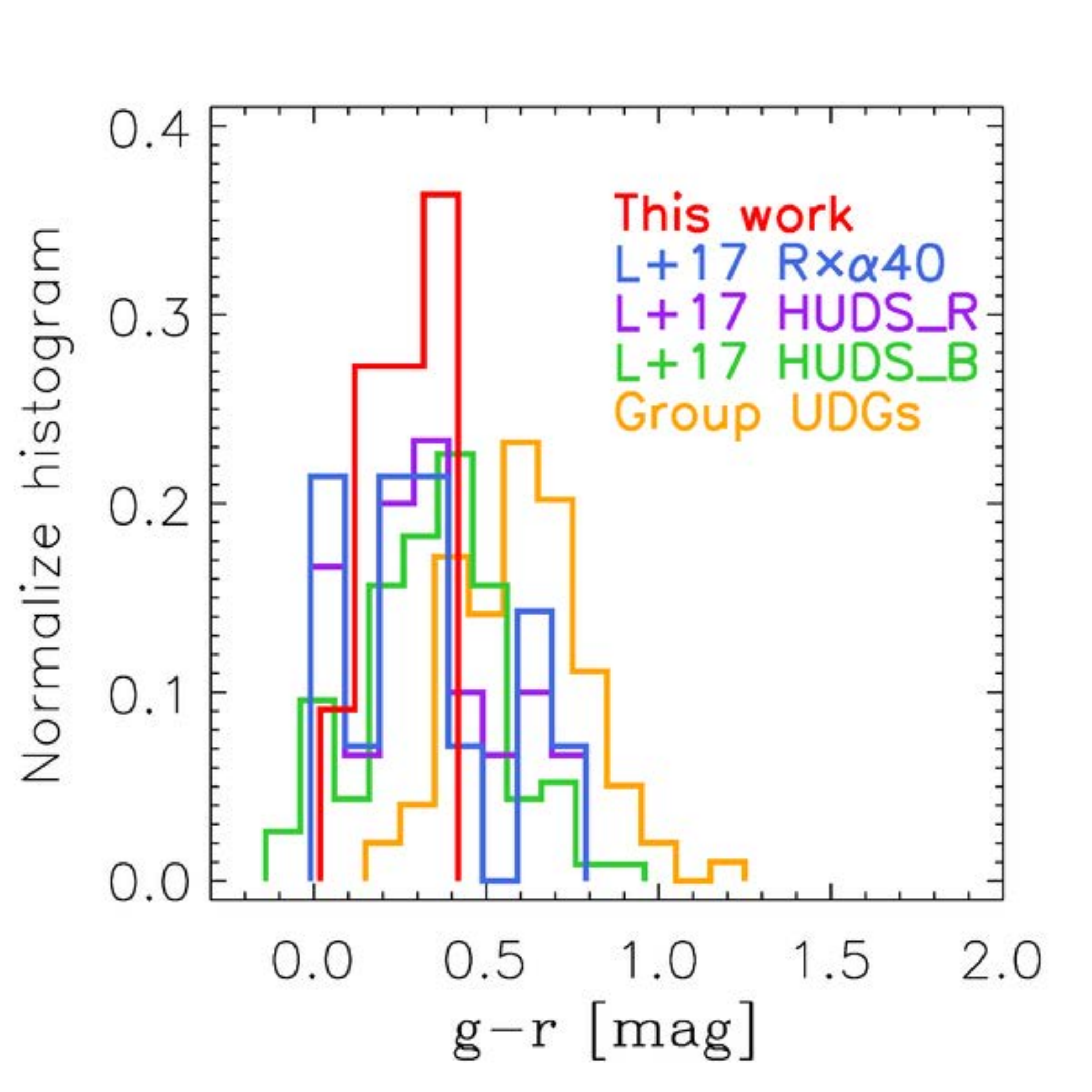}\\
\vspace{-0.1cm}
{\hspace{0.01in} (a) \hspace{1.5in} (b) \hspace{1.5in} (c) \hspace{1.5in} (d)}
\vspace{0.0cm}
\includegraphics[width=0.24\textwidth]{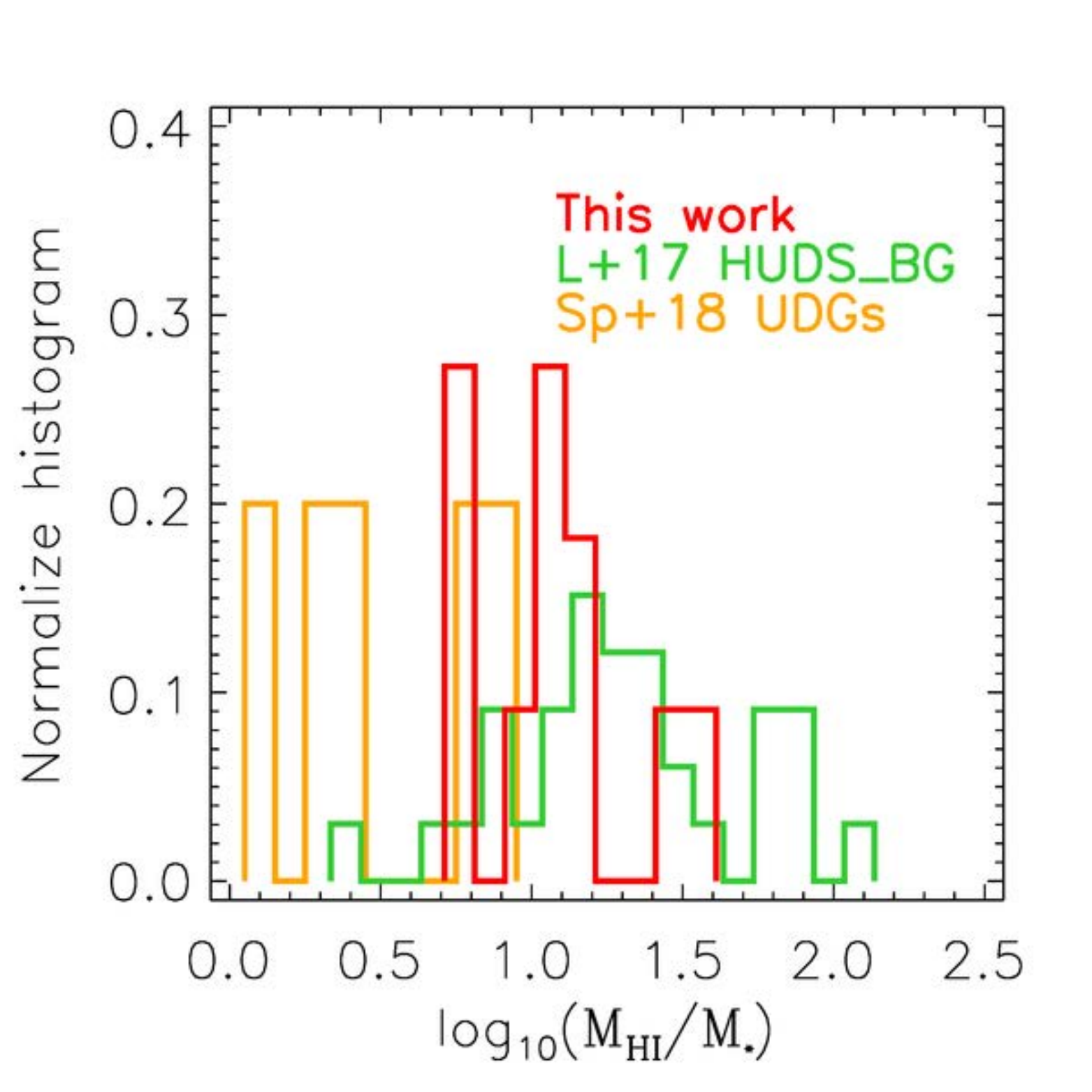}
\includegraphics[width=0.24\textwidth]{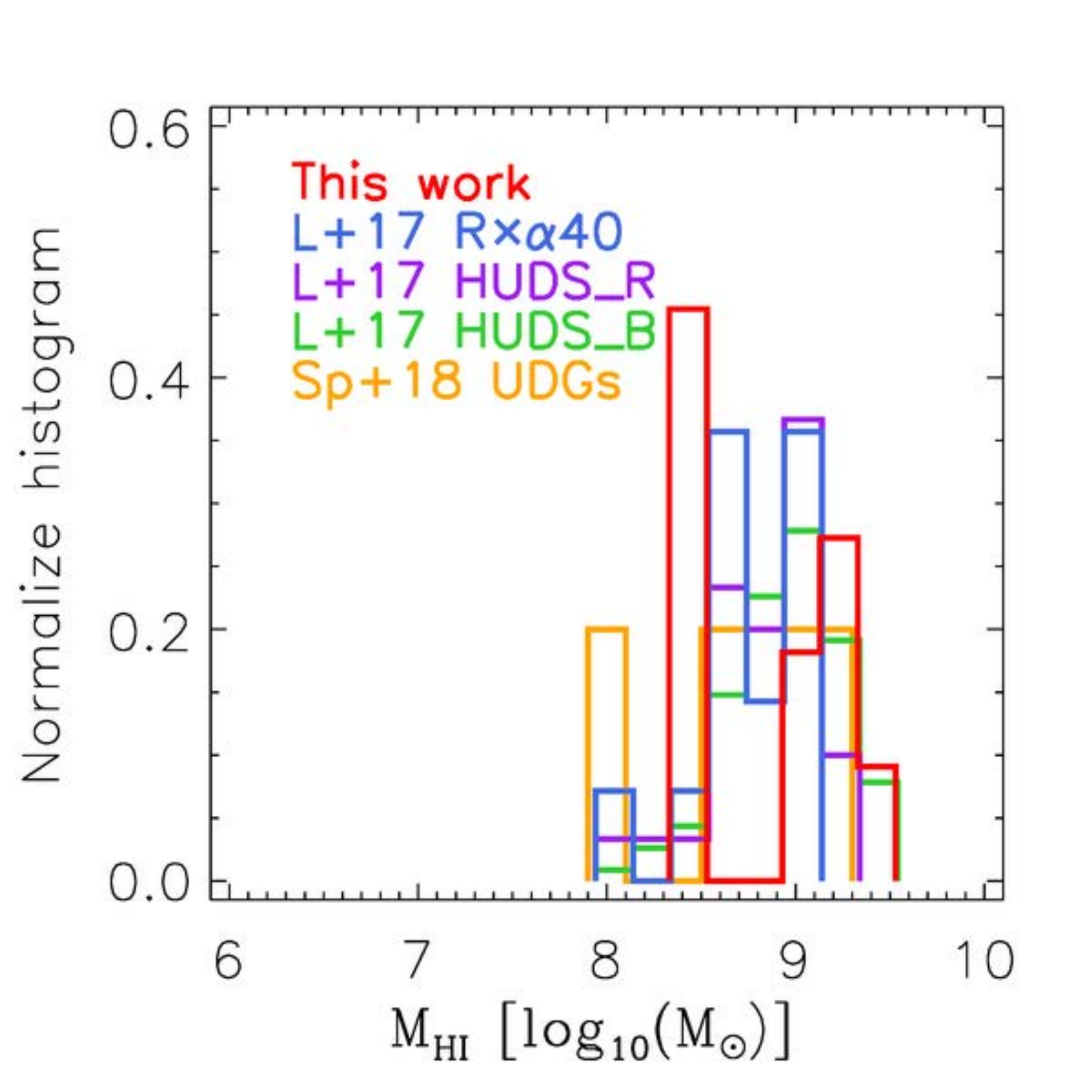}
\includegraphics[width=0.24\textwidth]{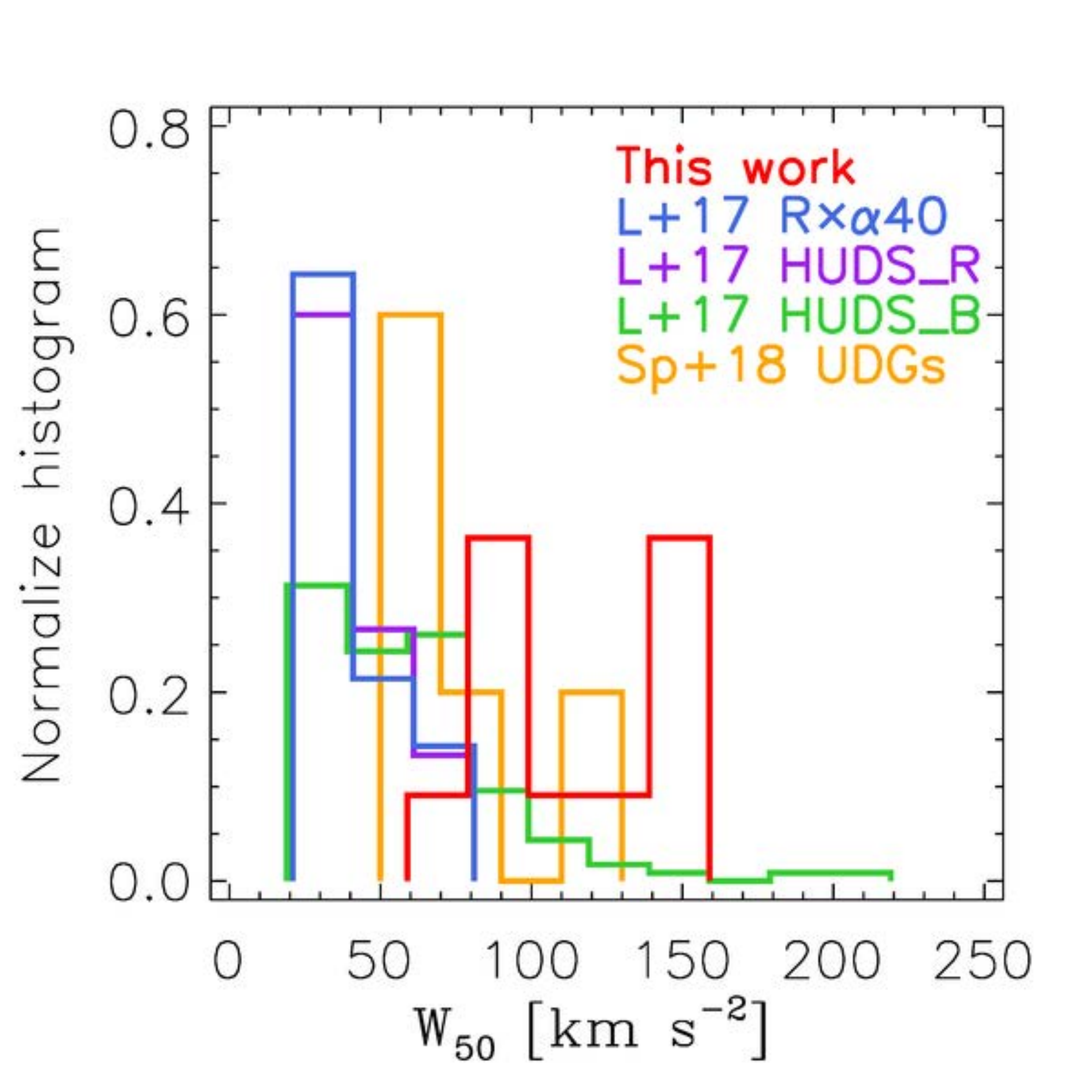}
\includegraphics[width=0.24\textwidth]{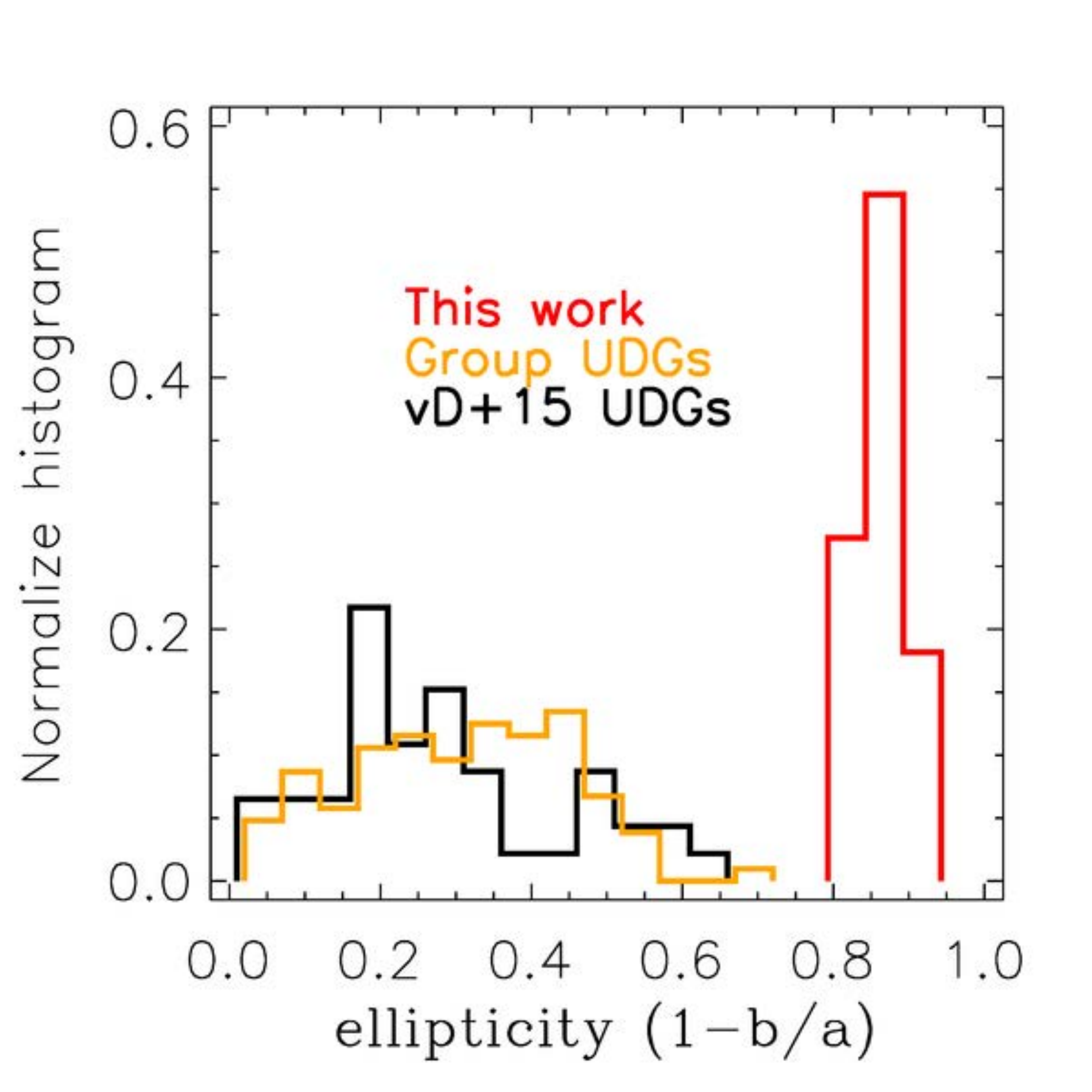}\\
\vspace{-0.1cm}
{\hspace{0.01in} (e) \hspace{1.5in} (f) \hspace{1.5in} (g) \hspace{1.5in} (h)}
\vspace{0.0cm}
\caption{Top: (a)-(d): are histograms showing optical properties ($M_{g}$, $r_{e}$, $\mu_{g,0}$, color $g-r$) of HUDS candidates compared with samples from the literature. Bottom: (e)-(h): are histogram comparisons of H{\sc{i}} relative properties (ratio of H{\sc{i}} to stellar mass, H{\sc{i}} mass and H{\sc{i}}-line velocity width $W_{50}$) and the ellipticity. Black represents UDGs of \citetalias{van.Dokkum..2015ApJ...798L..45V} in Coma Cluster. Orange means the total group UDGs found by \citet{Roman..2017MNRAS.468.4039R, Shi..2017ApJ.846.26, Muller..2018A&A...615A.105M, Merritt..2016ApJ...833..168M} or five blue UDGs which are studied in \citet{Spekkens..2018ApJ...855...28S}. Green and Purple correspond to HUDS\_B and HUDS\_R respectively, as selected from the $\alpha .70$ catalog \citepalias{Leisman..2017ApJ...842..133L}. Blue means the $\alpha .40$ part of HUDS\_R sample ($R\times \alpha .40$). Red indicates our 11 edge-on UDG candidates selected from $\alpha .40$ catalog. As this figure shows, the distributions of properties of our sources are similar to \citetalias{Leisman..2017ApJ...842..133L}'s sources, and have much larger ellipticity than other UDGs, which hint at high inclinations and thin shapes.}
\label{fig:properties}
\vspace{0.4cm}
\end{figure*}

\begin{table*}
\centering
\caption{Different stellar masses by using different methods}
\label{tab:diff-stellar-mass}
\setlength{\leftskip}{-30pt}
\resizebox{\hsize}{!}{
\begin{tabular}{cccccccccc}
\hline
\hline
 AGCNr
 & $lg(M_{H{\sc{I}}}/M_\odot)$
 & $lg(M_\star/M_\odot)$(Bell03)\footnote{\scriptsize stellar mass computed using Mass-to-light versus color relations of \citet{Bell..2003ApJ..149..289}}
 & $lg(M_\star/M_\odot)$(BC03)\footnote{\scriptsize stellar mass computed using Mass-to-light versus color relations in the Table A1. of \citet{Roediger..2015MNRAS.452.3209R}}
 & $lg(M_\star/M_\odot)$(FSPS)\footnote{\scriptsize stellar mass computed using Mass-to-light versus color relations in the Table A1. of \citet{Roediger..2015MNRAS.452.3209R}}
 & $lg(M_\star/M_\odot)$(Z09)\footnote{\scriptsize stellar mass computed using Mass-to-light versus color relations of \citet{Zibetti..2009MNRAS.400.1181Z}}
 & $M_{H{\sc{I}}}/M_\star$(Bell03)
 & $M_{H{\sc{I}}}/M_\star$(BC03)
 & $M_{H{\sc{I}}}/M_\star$(FSPS)
 & $M_{H{\sc{I}}}/M_\star$(Z09)\\

 &  &  &  &  &  &  &  &  & \\
\hline
 102276 & $9.10\pm 0.06$ & $8.69\pm 0.04$ & $8.38\pm 0.05$ & $8.51\pm 0.05$ & $8.35\pm 0.05$ & $ 2.58\pm 0.44$ & $ 5.21\pm  0.96$ & $ 3.89\pm 0.71$ & $ 5.68\pm  1.05$\\
 113816 & $9.06\pm 0.06$ & $8.34\pm 0.09$ & $8.06\pm 0.11$ & $8.18\pm 0.10$ & $8.02\pm 0.11$ & $ 5.26\pm 1.27$ & $ 9.99\pm  2.91$ & $ 7.58\pm 2.11$ & $10.87\pm  3.19$\\
 198457 & $9.20\pm 0.06$ & $8.51\pm 0.06$ & $8.10\pm 0.08$ & $8.25\pm 0.08$ & $8.06\pm 0.08$ & $ 4.95\pm 0.98$ & $12.53\pm  2.88$ & $ 8.83\pm 1.95$ & $13.81\pm  3.20$\\
 202262 & $8.35\pm 0.10$ & $7.71\pm 0.10$ & $7.33\pm 0.11$ & $7.48\pm 0.11$ & $7.29\pm 0.11$ & $ 4.37\pm 1.45$ & $10.38\pm  3.54$ & $ 7.44\pm 2.52$ & $11.40\pm  3.90$\\
 215226 & $8.33\pm 0.07$ & $7.97\pm 0.08$ & $7.66\pm 0.11$ & $7.79\pm 0.10$ & $7.62\pm 0.11$ & $ 2.28\pm 0.57$ & $ 4.71\pm  1.37$ & $ 3.50\pm 0.98$ & $ 5.14\pm  1.51$\\
 219242 & $9.26\pm 0.06$ & $8.71\pm 0.06$ & $8.38\pm 0.08$ & $8.51\pm 0.07$ & $8.34\pm 0.08$ & $ 3.56\pm 0.69$ & $ 7.67\pm  1.72$ & $ 5.64\pm 1.22$ & $ 8.39\pm  1.90$\\
 223141 & $9.49\pm 0.05$ & $8.73\pm 0.06$ & $8.38\pm 0.07$ & $8.52\pm 0.07$ & $8.34\pm 0.07$ & $ 5.76\pm 0.99$ & $12.78\pm  2.55$ & $ 9.33\pm 1.79$ & $13.99\pm  2.82$\\
 321194 & $8.48\pm 0.12$ & $8.10\pm 0.12$ & $7.80\pm 0.12$ & $7.92\pm 0.12$ & $7.76\pm 0.12$ & $ 2.42\pm 0.94$ & $ 4.84\pm  1.90$ & $ 3.63\pm 1.42$ & $ 5.27\pm  2.07$\\
 729579 & $8.50\pm 0.09$ & $7.44\pm 0.10$ & $7.06\pm 0.12$ & $7.21\pm 0.11$ & $7.02\pm 0.12$ & $11.43\pm 3.61$ & $27.32\pm  9.28$ & $19.54\pm 6.51$ & $30.02\pm 10.24$\\
 749223 & $8.49\pm 0.07$ & $7.50\pm 0.09$ & $7.02\pm 0.11$ & $7.19\pm 0.10$ & $6.98\pm 0.11$ & $ 9.88\pm 2.55$ & $29.42\pm  8.84$ & $19.86\pm 5.74$ & $32.67\pm  9.88$\\
 749493 & $9.16\pm 0.06$ & $8.46\pm 0.05$ & $7.11\pm 0.07$ & $8.25\pm 0.06$ & $8.07\pm 0.07$ & $ 5.04\pm 0.93$ & $11.13\pm  2.28$ & $ 8.13\pm 1.62$ & $12.19\pm  2.51$\\
\hline
\hline
\end{tabular}}
\vspace{-0.2cm}
\begin{flushleft}
\textbf{\footnotesize Notes.}\\
\end{flushleft}
\vspace{-0.5cm}
\end{table*}

\begin{figure}
\plotone{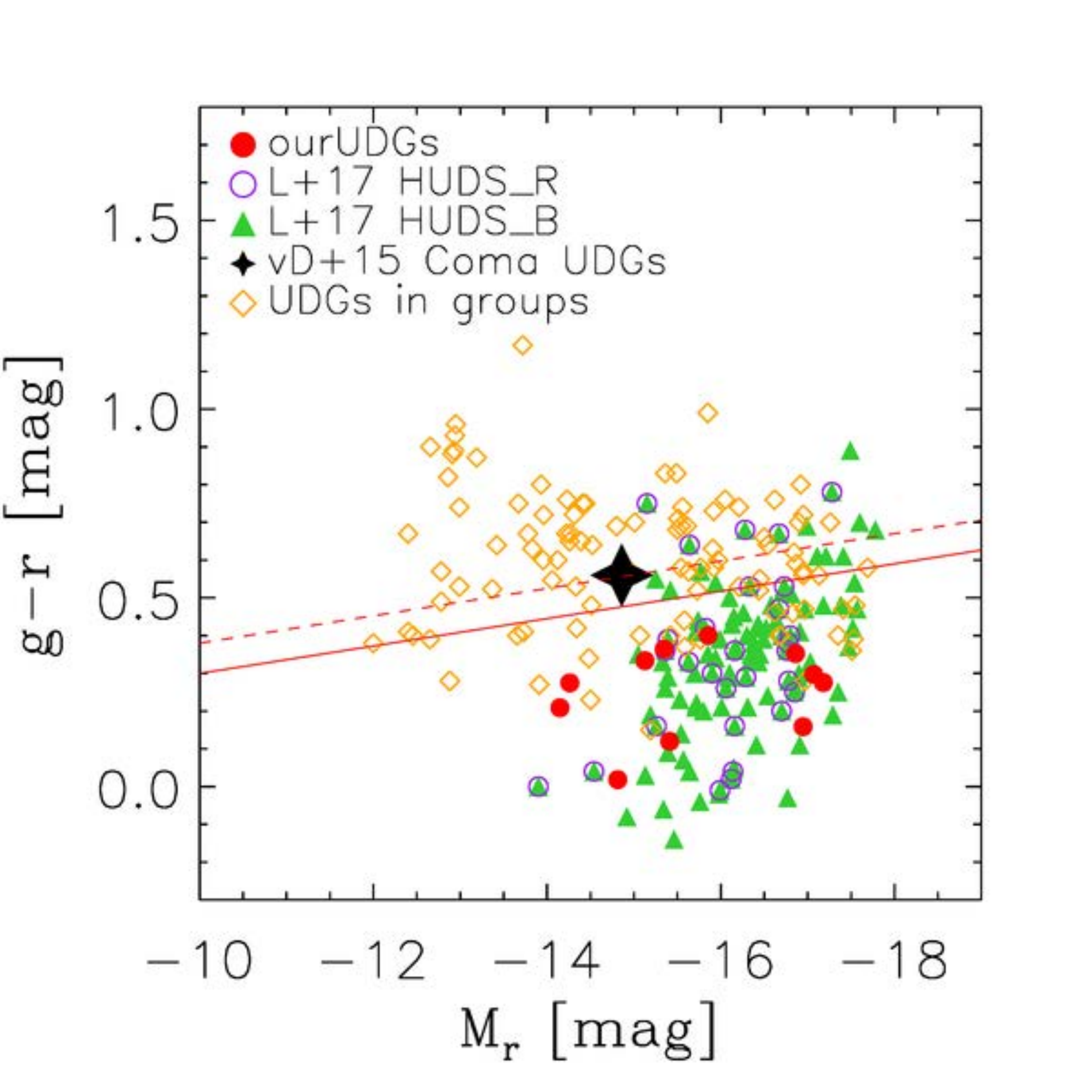}
\vspace{-0.4cm}
\caption{Color-absolute magnitude diagram of UDGs in fields and clusters. Green filled circles are our 11 edge-on HUDS candidates in $\alpha .40$, blue triangles are HUDS\_R of \citetalias{Leisman..2017ApJ...842..133L}, orange diamonds are the complete group of UDGs found by \citet{Roman..2017MNRAS.468.4039R, Shi..2017ApJ.846.26, Muller..2018A&A...615A.105M, Merritt..2016ApJ...833..168M} and the big black star is a rough mean value of UDGs in the Coma cluster \citepalias{van.Dokkum..2015ApJ...798L..45V}. The red solid line is the division line to separate the red and blue-sequence galaxies \citep{van_der_Burg.2015AA.577.A19} at a redshift of 0.013 (mean redshift of our HUDS) and the red dashed line is the fitting line of UDGs in clusters \citep{van_der_Burg.2016AA.590.A20}. This figure shows that all of our HUDS candidates and most HUDS\_R separate from UDGs in clusters.}
\label{fig:CMD}
\vspace{0.4cm}
\end{figure}

We list some major parameters of our 11 HUDS candidates in Table \ref{tab:params}. In this table, the AGC ID is the entry number in the Arecibo General Catalog (AGC) \citep{Haynes..2011AJ..142..170} and the R.A., Dec., distance, velocity width $W_{50}$ and H{\sc{I}} mass are from the $\alpha .40$ catalog. $r_{g,e}$ and $\mu_{g,0,edge}$ are derived by the GALFIT edge-on disk model fitting, while $\epsilon = 1-b/a$ and $b/a$ is derived from the GALFIT fitting with exponential disk model. $\mu_{g,0,face}$ is the face-on perspective central surface brightness corrected from $\mu_{g,0,edge}$ by Equation \ref{eq:correct-mu}. Stellar mass ($M_\star$) is calculated by using the method of \cite{Zibetti..2009MNRAS.400.1181Z} with $M_g$ and $g-r$ color.\\

Figure \ref{fig:properties} shows the distributions of general properties (central surface brightness $\mu _{g,0}$, effective radius $r_e$, absolute magnitude $M_g$, $g-r$ color, mass ratio $M_{H{\sc{I}}}/M_{\star}$, H{\sc{i}} mass, velocity width $W_{50}$ and ellipticity) of 11 HUDS candidates and other comparison UDGs. The compared UDGs are from the sample of isolated HUDS of $\alpha .70$ \citepalias{Leisman..2017ApJ...842..133L} and those in groups and clusters are from other works \citep[\citetalias{van.Dokkum..2015ApJ...798L..45V}][]{Merritt..2016ApJ...833..168M, Roman..2017MNRAS.468.4039R, Shi..2017ApJ.846.26, Muller..2018A&A...615A.105M}.\\

Our HUDS candidates are from the optical-H{\sc{i}} cross matches between SDSS DR7 and $\alpha.40$ \citep{Haynes..2011AJ..142..170}, while the \citetalias{Leisman..2017ApJ...842..133L} HUDS are from the cross matches between SDSS DR12 and $\alpha.70$. For comparison, we additionally compare the whole \citetalias{Leisman..2017ApJ...842..133L} HUDS\_R (purple lines) sample with those HUDS which are both included in the \citetalias{Leisman..2017ApJ...842..133L} HUDS\_R sample and optical-H{\sc{i}} cross matches between SDSS DR7 and $\alpha.40$ (blue lines). The HUDS\_R sample is selected by strict criteria of $\mu_{g,0} \geqslant 24$ mag sec$^{-2}$ and $r_{g,e} \geqslant 1.5$ kpc, while the HUDS\_B sample is selected by $\langle \mu_{r,eff}\rangle \geqslant 24$ mag sec$^{-2}$ and $r_{r,e} \geqslant 1.5$ kpc, broader than HUDS\_R. As Figure \ref{fig:properties} shows, there is no significant difference for the distribution of properties between the total HUDS\_R sample and its $\alpha .40$ part. It seems that despite a difference in amount, it will not take a significant difference in the distributions of properties between the $\alpha .40$ and $\alpha .70$ catalogs, as does SDSS DR7 and SDSS DR12.\\

\subsection{Optical properties}
The absolute magnitudes $M_g$ of our edge-on HUDS candidates range from $-14$ to $-17$ mag and the effective radii $r_e$ are from 1.5 to 5.67 kpc. Both are consistent with those of the HUDS\_R sample of \citetalias{Leisman..2017ApJ...842..133L}. Also, the edge-on HUDS candidates are very blue and the average color of $g-r$ is 0.26, even bluer than that of 0.344 for the HUDS\_R sample. But all the $g-r$ colors of edge-on HUDS are still in the range of the HUDS\_R sample. All these optical properties of edge-on HUDS candidates are similar to those of the HUDS\_R sample, and the blue colors of edge-on HUDS candidates also indicate that it is reasonable to neglect internal-extinction of edge-on HUDS.\\

However, we also make a comparison of ellipticity with other UDGs in Figure \ref{fig:properties}(h). As our candidates are all edge-on, the $g$-band ellipticity values of our HUDS candidates are all larger than $0.79$, and their mean value is $0.85$. The largest values of ellipticity for UDGs in the group \citep{Merritt..2016ApJ...833..168M, Roman..2017MNRAS.468.4039R, Shi..2017ApJ.846.26, Muller..2018A&A...615A.105M} and UDGs in the Coma cluster \citepalias{van.Dokkum..2015ApJ...798L..45V} are 0.7 and 0.62, while their mean values are 0.31 and 0.28, respectively. It is obvious that the ellipticity values of our HUDS candidates are really much larger than those of other UDGs. Because the $b/a$ or ellipticity of \citetalias{Leisman..2017ApJ...842..133L} were not available, we have not compared with \citetalias{Leisman..2017ApJ...842..133L}. However, we have cross-matched the HUDS\_B sample with our 1670 edge-on subsample, which are selected by $b/a \leqslant 0.3$ (mentioned in Section \ref{fig:galfit}), and no galaxy was matched. So, these edge-on HUDS we found have not been included in the HUDS\_B sample. By checking the DECaLS images of the HUDS sample, which are deeper than SDSS images, the morphologies of the HUDS\_B sample seem less likely to be high inclination galaxies.\\

\subsection{Stellar mass estimates}

Many methods exist for using optical colors to estimate stellar mass to light ratios \citep[e.g.][]{Bell..2003ApJ..149..289, Zibetti..2009MNRAS.400.1181Z, Taylor.2011MNRAS.418.1587T, Into.2013MNRAS.430.2715I, McGaugh.2014AJ.148.77M, Roediger..2015MNRAS.452.3209R}. But in some situations, there may be significant variance of stellar mass estimated by different methods, because of the dependence on SFH, initial mass function, and assumption of extinction \citep{Conroy.2013ARA&A..51..393C, McGaugh.2014AJ.148.77M, Herrmann.2016AJ.152.177H, Garcia-Benito.2019.AA.621A.120G}. \citet{Roman..2017MNRAS.468.4039R} uses the method provided by \citet{Roediger..2015MNRAS.452.3209R} to obtain a rough stellar mass. Also \citet{Roediger..2015MNRAS.452.3209R} provide two SPS model results, BC03 \citep{Bruzual..2003MNRAS.344.1000B} and FSPS \citep{Conroy.2009ApJ...699..486C}. This may be especially true for low surface brightness sources with extreme stellar populations. While \citetalias{Leisman..2017ApJ...842..133L} use the Z09 method for three sources with HI-synthesis observations, \citet{martinez..2016AJ..151..96M} give a comparison between \citet[][hereafter Z09]{Zibetti..2009MNRAS.400.1181Z} and \citet[][hereafter Bell03]{Bell..2003ApJ..149..289} for a single UDG in a much lower density environment, and conclude that Bell03 may be a factor of two higher than Z09.\\

So we estimate the stellar mass using all the methods mentioned above, Z09, Bell03, BC03 and FSPS, and compare the results of them to see whether the different methods would lead to a different conclusion. The results are displayed in Table \ref{tab:diff-stellar-mass}. From the comparison, stellar masses of these sources are lowest when applying Z09, and biggest when employing Bell03. It can be a factor of two higher than Z09, which is the same as discussed in \citet{martinez..2016AJ..151..96M}. But even if using Bell03, the ratios of $M_{H{\sc{I}}}$ and $M_\star$ also imply that our HUDS candidates are gas-rich galaxies. In addition, no matter which method has been chosen, the results of stellar mass are not beyond the range of stellar mass for the UDGs of \citet{Koda..2015ApJ...807L...2K}. For comparing our results with \citetalias{Leisman..2017ApJ...842..133L}, we finally adopt the Z09 method to calculate stellar mass, and our following discussion is also based on it. In this method,
\begin{equation}
\label{eq:mass-to-light-Z09}
lg(\frac{M/M_\odot}{L_\lambda /L_\odot}) = a_{\lambda}+(b_{\lambda}\times color)
\end{equation}
we use the $g$-band and $g-r$ color for calculating, and the $a_{g}$ and $b_{g}$ are $-1.030$ and 2.053 respectively. The stellar mass of our HUDS candidates ranges from $10^{6.98} M_{\sun}$ to $ 10^{8.35} M_{\sun}$, which is included in the range of \citetalias{Leisman..2017ApJ...842..133L}.\\

\subsection{HI properties}

All the H{\sc{i}} information about our HUDS candidates is derived from the $\alpha .40$, and we show in the Figure \ref{fig:properties}(e)-(g).\\

The distributions of stellar mass and H{\sc{i}} mass for our edge-on HUDS candidates demonstrate that they have H{\sc{i}} mass of $10^{8.33} M_{\sun} \sim 10^{9.49} M_{\sun}$, much larger than their stellar mass. Based on their stellar mass, they are optical dwarf galaxies, but they have medium mass of H{\sc{i}} gas ($7.7 \leqslant lg(M_{H{\sc{I}}}) \leqslant 9.5$), according to the classification of \cite{Huang.2012ApJ.143.133}.\\

As the Figure \ref{fig:properties}(e) demonstrates, our HUDS candidates are gas-rich galaxies, and their H{\sc{i}}-to-stellar mass ratios are from $5.14$ to $32.67$. The most gas rich edge-on HUDS is AGC 749223. Its H{\sc{i}} mass is $\sim$ 33 times of stellar mass. The comparison samples in Figure \ref{fig:properties}(e) are the HUDS\_BG sample (green line), which is cross-matched with the sample of \cite{Huang.2012ApJ.143.133}, and the five blue UDGs (orange line), which are found in Hickson Compact Groups (HCGs) by \cite{Roman..2017MNRAS.468.4039R}, are taken from H{\sc{i}} observations by GBT and VLA \citep{Spekkens..2018ApJ...855...28S}. From this figure, the mass ratio distributions of our HUDS candidates and HUDS\_BGs are similar, and are bigger than that of the five blue UDGs found in groups.\\

The velocity widths ($W_{50}$) of our edge-on HUDS candidates are from 59 to 150 km s$^{-1}$, much larger than those of the HUDS\_R sample, and it can be explained by the effect of inclination.\\

\subsection{Environment}

For examining the environment of our HUDS candidates, we also employ the criteria which is applied by \citetalias{Leisman..2017ApJ...842..133L} to check whether they are isolated sources. The criterion is to require the nearest galaxy, which has a heliocentric velocity within $500 km$ $s^{-1}$, has a projected separation farther than $350 kpc$. As \citetalias{Leisman..2017ApJ...842..133L} utilized a private catalog, the Arecibo General Catalog, we use both SDSS DR15 and the 100\% ALFALFA catalog instead. In this method, the first nearest neighbor only probes the local environment, and we have not examined the large scale environment of these HUDS candidates.\\

We find out that there are three HUDS (AGC 202262, AGC 215226 and AGC 729579) that have one nearby galaxy according to this criterion, and one HUDS (AGC 219242) has two neighbors. A nearby galaxy is not found to satisfy the criterion for the other candidates. So, our HUDS candidates are in low density environments, and 64\% of them are isolated galaxies. For considering this result clearly, we list all the nearby neighbors of the four HUDS candidates (AGC 202262, AGC 215226, AGC 219242 and AGC 729579), and the nearest galaxy of other HUDS candidates, which do not satisfy the criterion, in Table \ref{tab:nearest-galaxies}. As \citet{Du_Wei_2015AJ..149..199} compared the environment of HI selected non-edge-on low surface brightness galaxies with all the galaxies from the 40\% ALFALFA catalog, they are both more likely to reside in local low density environments. We also try to get the fraction of isolated galaxies from the 100\% ALFALFA catalog by the criteria mentioned in the section 2.1 of \citetalias{Leisman..2017ApJ...842..133L}, and the value is $\sim$65\%, which is similar to the proportion of isolated galaxies in our HUDS candidates. So similarly to all the galaxies from the entire ALFALFA catalog, our HUDS candidates selected from the $\alpha .40$ catalog also inhabit in local low density environment.\\

\begin{table}[h]
\caption{nearby galaxies of 11 HUDS candidates}
\label{tab:nearest-galaxies}
\setlength{\leftskip}{-15pt}
\resizebox{\hsize}{!}{
\begin{tabular}{cccccc}
\hline
\hline
 AGCNr\footnote{\scriptsize AGC ID of our 11 HUDS candidates.}
 & nearby galaxies \footnote{\scriptsize The ObjID or AGC ID of nearest galaxies found from SDSS DR15 or 100\% ALFALFA catalog.}
 & R.A.\footnote{\scriptsize Right ascension of the nearest galaxies.}
 & Dec. \footnote{\scriptsize Declination of the nearest galaxies.}
 & Separation \footnote{\scriptsize Projection separations between nearest galaxies and HUDS candidates.}
 & cz\footnote{\scriptsize Heliocentric velocities of the nearest galaxies.}\\
 & & (J2000) & (J2000) & (Mpc) & (km s$^{-1}$)\\
\hline
102276 & 1237680285992944294 & 00:51:18 & +30:26:56 & 5.485 & 5013\\
113816 & 1237679478017557091 & 01:42:43 & +21:32:11 & 7.47  & 4962\\
198457 & 1237658425159843953 & 09:57:21 & +07:11:19 & 1.833 & 6462\\
202262 & 5812                & 10:40:57 & +12:28:18 & 0.326 & 1008\\
215226 & 215197              & 11:44:44 & +15:01:40 & 0.226 & 3356\\
219242 & 213704              & 11:29:34 & +07:59:18 & 0.261 & 6274\\
219242 & 213830              & 11:30:01 & +07:39:38 & 0.322 & 6306\\
223141 & 1237658493357457493 & 12:40:17 & +10:31:07 & 0.802 & 6483\\
321194 & 1237672764977382079 & 23:06:33 & +24:22:57 & 0.729 & 1083\\
729579 & 212915              & 11:24:13 & +26:14:44 & 0.237 & 1482\\
749223 & 1237667447801577640 & 11:59:43 & +25:17:05 & 1.047 & 3435\\
749493 & 1237665442602418485 & 15:09:25 & +23:45:29 & 1.153 & 4278\\
\hline
\hline
\end{tabular}}
\vspace{-0.2cm}
\begin{flushleft}
\textbf{\footnotesize Notes.}\\
\end{flushleft}
\vspace{-0.5cm}
\end{table}

As both our edge-on HUDS candidates and the HUDS\_R sample are in low density environments, and have similar properties, it is very possible that our HUDS candidates are complement of the HUDS\_R sample. Compared with our edge-on HUDS candidates, though UDGs in clusters from \citetalias{van.Dokkum..2015ApJ...798L..45V} have a similar distribution of effective radius, they have fainter central surface brightness and lower absolute magnitudes (Figure \ref{fig:properties}(a)-(c)). Also, UDGs in clusters have a red color of $g-r$ $\sim$ 0.6 \citep{van_der_Burg.2016AA.590.A20} on average. Therefore, UDGs in the field and in clusters are quite different populations.\\

\section{Discussion}
\label{sec:discussion}
\subsection{Sample Uncertainty}
While current measurements result in a sample of 11 edge-on candidate HUDS, uncertainties in the measured parameters may have some impact on the size of the sample. If considering the measurement errors in Table \ref{tab:params}, there may be three objects that do not satisfy the criteria of $\mu_{g,0} \geqslant 24$. As for the internal extinction in Section 3.3, if we correct the difference of $0.02$ between edge-on and face-on galaxies, 10 out of our 11 HUDS candidates will still satisfy our definition of UDGs; if we consider the standard deviation of edge-on internal extinction $0.08$,  six out of our 11 HUDS candidates will still meet our definition of UDGs.\\

\subsection{Inclination Selection effect}

With an equal total luminosity, like $M_g$, the observed central surface brightness $\mu_{g,0}$ of an edge-on disk-like galaxy is brighter than that of a face-on galaxy. If a face-on galaxy has $\mu_{g,0,face}=24$ mag arcsec$^{-2}$ and $h_s/r_s=0.3$, its observed central surface brightness will become $\mu_{g,0,edge}=22.69$ when it turns to an edge-on perspective, and then it will be missed by the criteria for UDG. Also, for the observed galaxies which satisfy the $\mu_{g,0}$ criterion of $\mu_{g,0} \geqslant 24$ mag arcsec$^{-2}$, the edge-on galaxies should have much lower luminosity than face-on galaxies, and also have less morphological characteristics that appear in images. This makes the edge-on galaxies become more difficult to detect. Therefore, it may lose many edge-on galaxies, which have a disk galaxy profile. Just like our edge-on HUDS candidates, they have not been found in \citetalias{Leisman..2017ApJ...842..133L}. So it is very possible that there is a selection effect for searching disk-like UDGs, like Figure \ref{fig:properties}(h) illustrates. Maybe that is why there is not a sufficient edge-on UDG sub-population in \citetalias{Leisman..2017ApJ...842..133L},  \citetalias{van.Dokkum..2015ApJ...798L..45V} and many other works. But, this inclination selection effect mainly influences exponential disk profile galaxies, and would decrease with the increase of S\'{e}rsic index, and can be neglected for irregular galaxies.\\

\subsection{Mechanism of UDG candidates}

The UDGs in clusters could be explained by failed $L_{\star}$ galaxies \citepalias{van.Dokkum..2015ApJ...798L..45V}, regular or irregular dwarf galaxies \citep{Beasley..2016ApJ.819.L20, Beasley..2016ApJ.830.23, Amorisco..2016mnras..459..L51} and tidal dwarfs \citep{van.Dokkum.2018Nature}. However, the low density environments and regular morphology of these edge-on HUDS candidates selected from $\alpha .40$ indicate that they are less likely to be tidal dwarfs. Also, the dwarf irregular galaxy is not a good explanation for our sources, as most of our HUDS candidates are disk-like galaxies. Although our HUDS candidates have an H{\sc{i}} mass of $lg(M_{HI}/M_{\sun})$ from 8.33 to 9.5, the low surface density environment and low star formation efficiency make them difficult to transform H{\sc{i}} gas to stars. Their slow evolution would be the key reason that they only have small stellar mass like dwarf galaxies. Even if all the H{\sc{i}} gas can be transformed to stars, their stellar masses are still far less than the stellar mass of $\sim 5 \times 10^{10}$ of $L_{\star}$ galaxies. Our edge-on HUDS candidates are medium mass disk-like UDGs and cannot be explained by failed $L_{\star}$ galaxies. \\

Figure \ref{fig:CMD} shows the color-absolute magnitude diagram of UDGs in fields, groups and clusters. The red solid line is the division line to separate the red-sequence and blue-cloud galaxies \citep{van_der_Burg.2015AA.577.A19}. We adopt the redshift of 0.013 for drawing the division line, which is the mean redshift of our HUDS. All of our edge-on HUDS are blue galaxies and located below the division line. Similarly, most of the HUDS\_R cases are also located at the same region, except some red ones. The red dashed line is the line with an offset of 0.08 mag from the division line \citep{van_der_Burg.2016AA.590.A20}.\\

The $g-r$ colors of the UDGs in the Coma Cluster, which have been detected by \citetalias{van.Dokkum..2015ApJ...798L..45V}, are not provided. But as pointed out by \citetalias{van.Dokkum..2015ApJ...798L..45V}, the average color of their UDGs is consistent with low metallicity or young age passively evolving stellar populations, such as the age of $7Gyr$ and the metallicity of $[Fe/H]=-1.4$ predicted by \cite{Conroy..2009ApJ...699..486C}. So, we estimate the mean value of color $g-r$ by using the template corresponding with the BC03 model \citep{Bruzual..2003MNRAS.344.1000B} with the age of $7Gyr$ and the metallicity of $[Fe/H]=-1.4$. Then we obtain the color $g-r \sim 0.56$ and plot this value by a big black star in Figure 5. Similar to normal galaxies in the color-absolute magnitude diagram, the UDGs can also be divided into red UDGs and blue UDGs. Like previous works about evolution of galaxies in clusters \citep{Mayer..2001ApJ...559..754M, Kazantzidis..2011ApJ...726...98K, Bosch..2008MNRAS.387...79V, Amorisco..2016mnras..459..L51, Di.Cintio.2017MNRAS.466L.1D, Papastergis.2017A&A.601.L10, Chan..2018MNRAS.478..906C, Carleton..2018arXiv}, a possible connection may exist between field UDGs with the UDGs in clusters, and fall into the galaxy clusters for some reasons, like through ram pressure stripping and tidal stirring.\\

\section{Summary} \label{sec:summary}

Based on the ALFALFA 40\% catalog and SDSS-DR7 $g$-/$r$- band images, we have selected 11 edge-on HUDS candidates after correcting the edge-on central surface brightness to the face-on one. All these edge-on HUDS candidates are very blue and H{\sc{i}}-bearing, mostly isolated from nearby neighbors, and have properties consistent with those HUDS from ALFALFA 70\% catalog \citepalias{Leisman..2017ApJ...842..133L}. \\

In this paper, we focus on the detection of edge-on disk-like UDGs. These sources may be excluded by the criteria used to find UDGs if authors do not correct their central surface brightness values, but they can be easier to detect with an equal total magnitude, and they are good for studying their rotational velocity. So, we can improve the population of UDGs at the high inclination tail by using the correction of $\mu_{g,0}$. Also, the consistency of properties between our HUDS candidates and HUDS detected from \citepalias{Leisman..2017ApJ...842..133L} demonstrates that our approach is an efficient and reasonable way to select UDGs from edge-on systems.\\

\textbf{Acknowledgment} This project is supported by the National Natural Science Foundation of China (Grant No. 11733006), the National Key R$\&$D Program of China (No. 2017YFA0402704), the National Natural Science Foundation of China (Grant No. 11403037), and the Key Laboratory of Optical Astronomy, National Astronomical Observatories, Chinese Academy of Sciences. The authors sincerely thank the Sloan Digital Sky Survey project for providing their fpC-images for processing again, and thank the ALFALFA project for providing the catalog of matched 40\% and SDSS DR7 data. The authors also thank Huijie Hu for his help.\\

\bibliography{UDG}
%\end{CJK*}
\end{document}